\documentclass[a4paper,english,aps,twocolumn,floatfix,amsfonts,amssymb,superscriptaddress]{revtex4}
\usepackage{graphicx}
\usepackage{epstopdf} 
\usepackage{dcolumn}
\usepackage{bm}
\usepackage{bbm,bbold}
\usepackage{color}
\usepackage{xcolor, soul}
\sethlcolor{green}
\usepackage{amsfonts}
\usepackage{amsmath}
\usepackage{mdframed}
\usepackage[normalem]{ulem}
\usepackage{mathrsfs}   
\usepackage{fdsymbol}

\usepackage[percent]{overpic}
\usepackage[colorlinks=true,citecolor=blue]{hyperref}
\hypersetup{colorlinks=true,citecolor=blue,linkcolor=blue,urlcolor=blue}
\DeclareRobustCommand{\rchi}{{\mathpalette\irchi\relax}}
\newcommand{\irchi}[2]{\raisebox{\depth}{$#1\chi$}} 

 \newcommand\sbullet[1][.5]{\mathbin{\vcenter{\hbox{\scalebox{#1}{$\bullet$}}}}}

\begin{document}
\title{Dominant Role of Two-Photon Vertex in Nonlinear Response of Dirac Materials}
\author{Habib Rostami}
\email{habib.rostami@su.se}
\affiliation{Nordita, KTH Royal Institute of Technology and Stockholm University, Roslagstullsbacken 23, SE-106 91 Stockholm, Sweden}
\author{Emmanuele Cappelluti} 
\affiliation{Istituto di Struttura della Materia (ISM), CNR, 34149 Trieste, Italy} 
\begin{abstract}
Using a conserving Baym-Kadanoff approach,
we present a fully compelling theory of nonlinear dc response of a Dirac system 
to electric fields in the presence of disorder scattering.
We show that the nonlinear terms are strikingly ruled by the appearance of
a dominant two-photon vertex which is absent at the bare level and finite even in the weak-coupling limit. 
Such two-photon vertex self-generation highlights the crucial role of the frequency and field dependence of the scattering rates in the nonlinear regime. Our study reveals a novel many-body mechanism in the nonlinear response of Dirac materials whose effects are predicted to be observable. 
\end{abstract}
\maketitle

Due to their linear dispersion, $\epsilon_{\bf k} \sim |{\bf k}|$, and to the
underlying chiral structure, Dirac materials show a variety of exotic features
that makes them a versatile platform for theoretical investigations of new physics
and for application purposes.
Despite the complex physics, many properties of these materials
are often rationalized using concepts of non-interacting particle or semi-classical model \cite{Neto_graphene_rev_2009,Wehling_DMrev_2014,Armitage_DMrev_2018,katsnelson_book}.
For instance, a standard transport model is conventionally applied for the dc conductivity
in highly-doped graphene ({\em Boltzmann} regime), where the mobility is evaluated at the non-interacting level,
and the interactions enter only through the effective parameter known as
transport scattering rate $\Gamma_{\rm tr}$ \cite{dassarma_rmp}.
At odds with the above scenario, there is a wide consensus that the {\em quantum} regime
(low-energy transitions in undoped Dirac model) is much more complex and
it might be significantly affected by many-body effects \cite{peres_rmp}.

The nonlinear electromagnetic response of Dirac materials has attracted recently a considerable interest in two
\cite{Hendry_prl_2010,Zhang_optlett_2012, Kumar_prb_2013,Woodward_2DM_2016, Soavi_natnano_2018,Hafez_nature_2018,Soavi_acsphotonics_2019,Ma_XU_2019} and three dimensions \cite{Ma_2017,deJuan_2017,Rostami_prb_2018,Ma_2019,Cheng_prl_2020,cheng_2020}.
Widely investigated are the nonlinear optical properties and in particular
the appearing of four-wave mixing, nonlinear Kerr effect, second and third-harmonic-generation in single-layer graphene,
with remarkable technological interest 
\cite{Mikhailov_EPL_2007,Cheng_2014_njp,Cheng_prb_2015,Mikhailov_prb_2016,Rostami_prb_2016,Rostami_prb_2017a,Mikhailov_prb_2019,Principi_prb_2019,Hafez_review_thz,Rostami_prr_2020}. 
Peculiar of Dirac material is, due to the linear dispersion,
the absence of the bare two-photon-electron coupling,
which should give rise to the so-called {\it diamagnetic} term.
The lack of such term prompts several widely debated issues,
as the validity of optical sum rules \cite{Goldman_1982,Cenni_2001,Sabio_2008}.
Most of the theoretical descriptions of nonlinear effects rely
at the moment upon non-interacting analyses, or semiclassical approaches \cite{Cheng_prb_2015,Mikhailov_prb_2016,Parker_2019,Shen_book,Boyd_book} where, in a similar way as in the Boltzmann theory of linear response,
the dominant transition processes resemble the ones of the non-interacting case
and the scattering sources are accounted through effective parameters
as the scattering rate $\Gamma$
(or equivalently through the mean-free path $l$, the lifetime $\tau$, etc.).

In this Letter, we show that the a compelling analysis
of the many-body physics, beyond the semi-classical approaches,
can drastically change the above scenario, pointing out
that different physical processes can be responsible
for the relevant properties of the nonlinear dc transport.
Analyzing the case of disorder scattering as a basilar benchmark example,
we show how non-conserving phenomenological models
of scattering intrinsically fail and high-order vertex processes must
properly taken into account.
More in particular, we show that, despite the bare diamagnetic
two-photon vertex (TPV) being null in Dirac materials,
the many-body renormalized TPV is finite
and relevant and it can play a dominant role.
Our results, besides providing a consistent framework
for a proper analysis of nonlinear transport and optical response
in realistically interacting Dirac materials, open
novel perspectives for understanding and predicting
new functional properties of these complex promising systems.

\begin{figure*}[t]
\centering
\includegraphics[width=150mm]{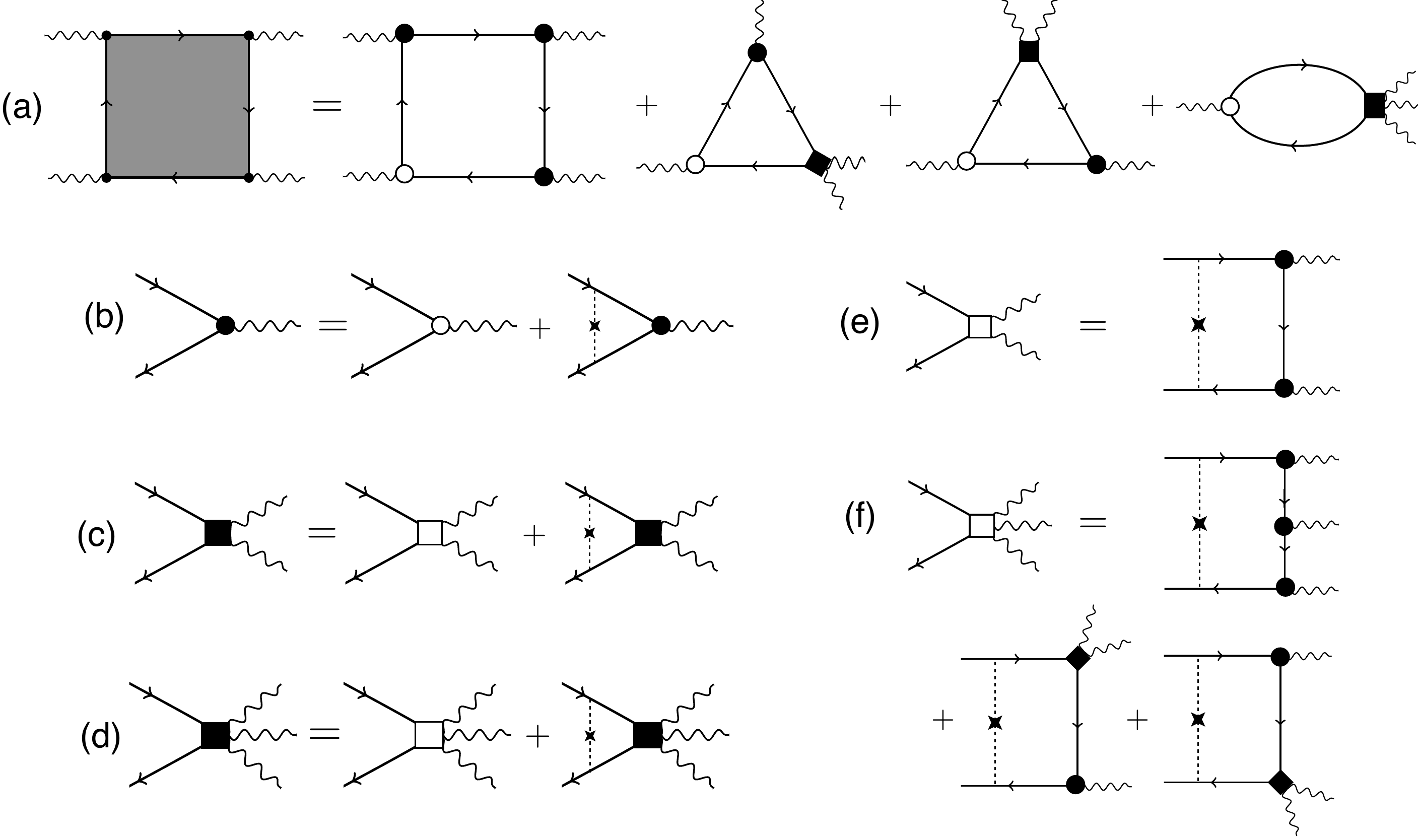}
\caption{{\bf Diagrammatic representation of nonlinear e.m. response in Dirac materials.} (a) nonlinear response expressed in terms
of {\em renormalized} one-, two- and three-photon vertices; (b)-(d) self-consistent Bethe-Salpeter eqautions
for  one-, two- and three-photon vertices; (e)-(f) many-body definition of {\em unrenormalized} two- and three-photon vertices in terms of lower order vertices.
Note that solid and wavy lines stand for fermion propagator and external photons, respectively. 
Dashed line indicate impurity interaction line. 
Void and filled circle stand for bare and renormalized one-photon vertex, respectively. 
Void and filled square stand for unrenormalized and renormalized two and three-photon vertices. 
}
\label{fig:chi3}
\end{figure*}

We consider the two-dimensional (2D) Dirac Hamiltonian
$\hat {\cal H}_{\bf k} =  \hbar v \hat{\bm \sigma}\cdot {\bf k}-\mu_0\hat I$,
where $\mu_0$ is the bare chemical potential ruling the charge doping.
For realistic purposes we consider the paradigmatic case of graphene \cite{notegraphene}
where $\hat{\bm \sigma} = (\tau\hat\sigma_x,\hat\sigma_y)$,where $\hat{\sigma}_i$
stand for the Pauli matrices in the spinor space, and
$\tau=\pm$ stands for valley index in the Brillouin zone of graphene.
In the dipole approximation the light-matter interaction can be modelled by applying
the minimal coupling transformation 
$\hbar{\bf k} \to \hbar{\bf k} + e{\bf A}(t)$ where ${\bf A}(t)$ stands for an external vector potential.
The corresponding electric field is given by ${\bf E}(t)=-\partial_t {\bf A}(t)$. 
Due to the linear dispersion, the electron-photon coupling does not present a diamagnetic
(two-photon) bare term but only the linear coupling: 
\begin{equation}
{\cal H}_{\rm light-matter}= \hbar ev \int d{\bf r}~\hat\psi^\dagger({\bf r}) \hat {\bm \sigma}\cdot{\bf A}(t)\hat\psi({\bf r})~.
\end{equation}
Without the loss of generality, we assume an electric field along the $y$ axis.
As scattering source
we consider long-range impurity centers with
standard Born impurity correlations 
 \cite{Shon_JPSJ_1998,Ando_JPSJ_2002,Rostami_prb_2017b}. 
Within this framework we can write the Born impurity self-energy in the complex frequency space:
$\hat \Sigma(z) = \gamma_{\rm imp} \sum_{\bf k} \hat G({\bf k},z)$
where the Green's function follows
$\hat G({\bf k},z) =[z - \hat {\cal H}_{\bf k}-\hat \Sigma(z)]^{-1}$.
For isotropic scattering we get a diagonal self-energy in the spinor basis as $\hat \Sigma (z) = \Sigma(z) \hat I$.
It is well known that under these conditions the impurity self-energy, as well
the Coulomb and other scattering ones, depends intrinsically on
the ultraviolet energy cut-off $W$ 
representing the range of validity of the Dirac model.
In order to provide a conserving approach,
this issue needs to be cured by means of a proper
regularization \cite{Leibbrandt_rmp_1975,Peskin}. As detailed in the Supplementary Material (SM), we employ standard dimensional regularization
leading to:
\begin{equation}
\label{eq:self}
\Sigma(z) = -U S(z) \ln[-W^2/S^2(z)],
\end{equation}
where $S(z) = z+\mu_0-\Sigma(z)$,
and $U$  is a dimensionless parameter characterizing the strength of impurity scattering \cite{Shon_JPSJ_1998,Ando_JPSJ_2002,Rostami_prb_2017b}.

Conserving approaches, based for instance on a Baym-Kadanoff derivation \cite{Baym_Kadanoff_prb_1961,Baym_Kadanoff_book},
are fundamental in theoretical physics to ensure that compelling results are obtained.
This aim is particularly important in nonlinear response since an arbitrary selection
of diagrams can easily lead to spurious conclusions.
The choice of the vector-potential gauge, within the
paradigmatic Born impurity scattering we consider here, permits us
an exact derivation of self-consistent equations (see SM for details \cite{SM}) for all the high order processes
relevant in the third-order response function which is the leading nonlinear term in centrosymmetric Dirac materials. The diagrammatic expression of the third-order response function
is provided in Fig.~\ref{fig:chi3} where,
roughly speaking, empty symbols represent $n$-photon vertices ($n=2,3$)
expressed in terms of the renormalized lower-order vertices (Fig.~\ref{fig:chi3}e,f),
whereas filled symbols represent the solution
of a Bethe-Salpeter-like (BS) self-consistent resummation
for a given $n$-photon vertex (Fig.~\ref{fig:chi3}b-d).
Leaving aside the complexity of the self-consistent set of equations,
few relevant things are worth to be underlined here.
First of all, we notice that an effective multi-photon coupling
is induced by the disorder scattering source even if it is absent
in the Hamiltonian at the bare (non-interacting) level
(Fig.~\ref{fig:chi3}e,f).
Second, that the relevance of each $n$-photon vertex
is largely governed by the BS many-body resummation
as depicted in Fig.~\ref{fig:chi3}b-d.
This might lead to a reduction (screening) or to an enhancement
of different multi-photon scattering depending on the Pauli structure
of the corresponding photon vertex, as we discuss more extensively
later.

The diagrammatic expressions in Fig.~\ref{fig:chi3} represent in full generality
the optical frequency-dependent third-order response function in Dirac materials, including 
third-harmonic  generation, four-wave mixing, etc.
For a generic interaction, the effective solution of such coupled equations 
on the real-frequency axis is a formidable task that does not allow for a practical solution.
The focus on the isotropic disorder scattering is on the other hand particularly suitable to
investigate many-body effects in nonlinear electromagnetic response
since it allows for a set of equations in the Matsubara space
which can be generalized in a rigorous way on the real frequency axis,
using the well-known procedure of multiple branch cuts 
in the complex frequency space.
The derivation is lengthly and cumbersome but compelling
and it is summarized in the SM \cite{SM}.
We consider first the dc transport limit.
Without loss of generality, it is possible to express the linear
and the third-order dc conductivity in terms of two  dimensionless quantities:
\begin{eqnarray}
\label{eq:sigma1n}
\sigma^{(1)}_{\rm dc}
&=&
\sigma_0 f_1\left(\frac{\mu}{\Gamma(\mu)};U \right),
\\
\label{eq:sigma3n}
\sigma^{(3)}_{\rm dc}
&=& 
\frac{\sigma_0}{{E}^2_{0}} \left[\frac{t_0}{\Gamma(\mu)} \right]^4
f_3\left(\frac{\mu}{\Gamma(\mu)};U\right),
\end{eqnarray}
where $\mu$ is the effective chemical potential $\mu=\mu_0-\mbox{Re}\Sigma(\omega=0)$
and $\Gamma(\mu)$ the scattering rate $\Gamma(\mu)=-\mbox{Im}\Sigma(\omega=0)$,
$\sigma_0 \propto e^2/\hbar$ the universal conductivity unit 
and ${E}_{0}\propto t_0/e a$ is a characteristic electric field scale
determined by inter-atomic hopping energy $t_0$ and by the lattice constant $a$ \cite{notegraphene}.

It is worth to stress again that Eqs.~(\ref{eq:sigma1n})-(\ref{eq:sigma3n})
are tied together since they must descend in a compelling way from
a common approximation for the self-energy.
Heretofore, although many approaches for the self-energy have been discussed for
the linear response, the third-order response has been analyzed only
in the simplistic case of a phenomenological constant scattering rate $\Gamma(\mu)=\Gamma$.
Since such phenomenological self-energy does not depend on the
applied external field, the third-order response function reduces to the first
``square'' diagram of Fig.~\ref{fig:chi3}a dropping all the vertex renormalization processes,
i.e. replacing the filled circles with empty ones (= bare electron-photon coupling).
A similar scheme can as well be employed for the linear response.
Under these ultra-simplified conditions, one can see that 
the linear and third-order dc transport depend uniquely on the
semiclassical parameter $x=\mu/\Gamma(\mu)$, i.e.
$f_1(x;y)=f_1(x)$, $f_3(x;y)=f_3(x)$.
An analytical expression for
the functions $f_1(x)$, $f_3(x)$ is obtained in the SM \cite{SM}.
In particular, in the Boltzmann regime one gets results 
$f_1(\infty) \approx 2 \mu/\pi\Gamma$,
$f_3(\infty) \approx -3\pi \Gamma/32\mu$,
implying that nonlinear effects lead
to a {\em reduction} of the dc conductivity in the Boltzmann regime.
A similar analysis is performed in the quantum regime, giving \cite{SM}
$f_1(0) = 8/\pi^2$, $f_3(0) =2/5$, meaning that
nonlinear effects should yield
an {\em enhancement} of the dc conductivity in the quantum regime.

The above predictions, based on the phenomenological model
of a constant scattering rate $\Gamma$, are challenged when
many-body effects are computed in a compelling conserving scheme.
\begin{figure}[t]
\includegraphics[width=90mm]{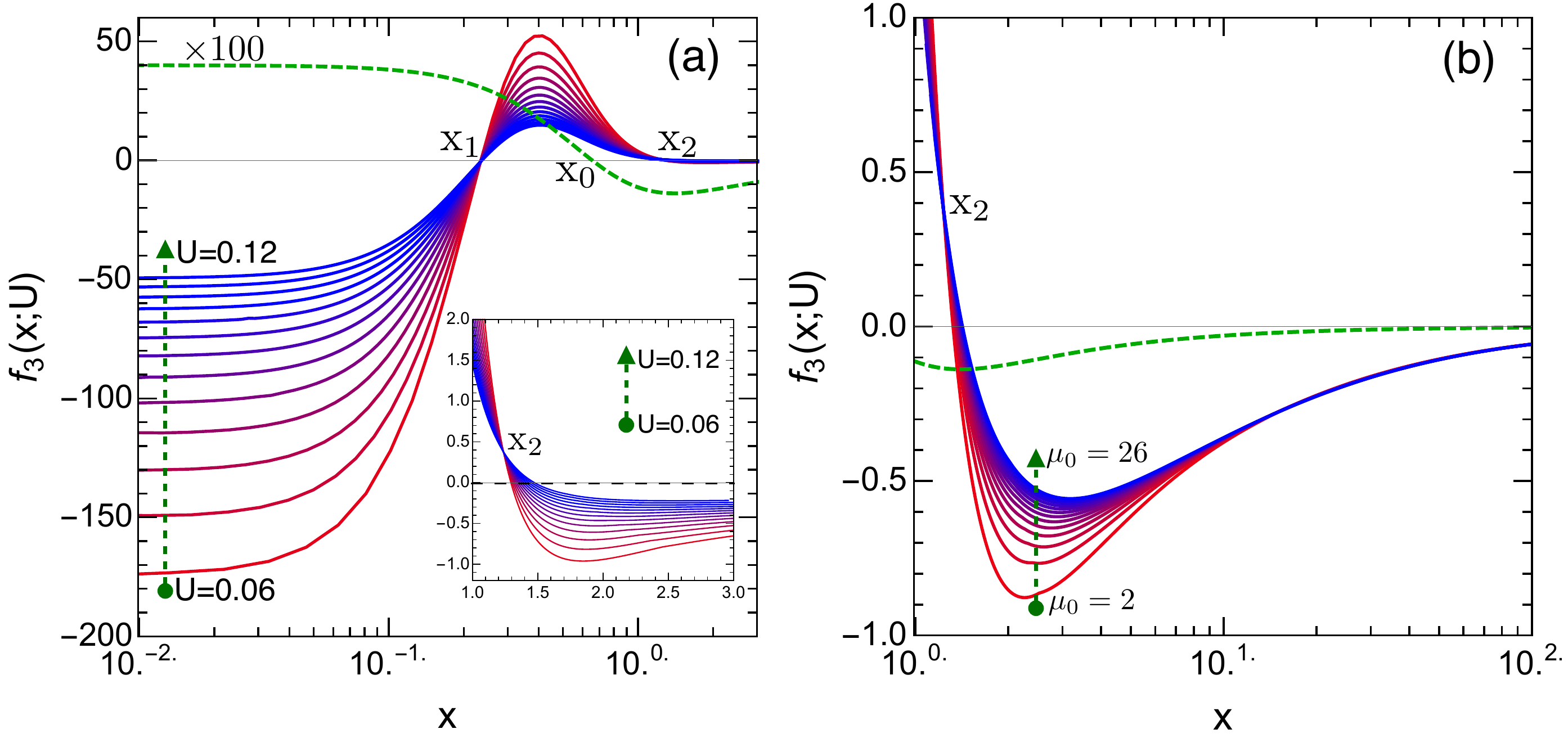}
\caption{
Characteristic nonlinear dc transport function $f_3$
as a function of the dimensionless parameter $x=\mu/\Gamma(\mu)$
for the many-body conserving scheme. Also shown is $f_3$
for the phenomenological model (dashed line).
Curves of $f_3$ vs. $x$ are plotted for different $U$'s in the quantum regime (panel a),
and for different $\mu_0$'s in the Boltzmann regime  in THz unit (panel b).
}
\label{f-vs-x}
\end{figure}
In Fig. \ref{f-vs-x} we show the characteristic dc transport~\cite{notethg} function $f_3$
as a function of the semiclassical parameter $x=\mu/\Gamma(\mu)$,
from the extreme quantum limit ($x \ll 1$) to the Boltzmann regime ($x \gg1$).
From the computational point of view, since the presence of a finite
cut-off energy scale $W$, the quantum limit can be conveniently investigated
by fixing $U$ and varying $\mu$, whereas the Boltzmann regime
is more easily spanned by fixing $\mu$ and varying $U$.
Let's discuss first the Boltzmann regime (Fig. \ref{f-vs-x}b).
We notice that a compelling many-body analysis recovers qualitatively
(but with an increase in the magnitude of factor 10)
the predictions of the phenomenological constant-$\Gamma$ model
with $f_3(x,y)\approx f_3(x)\propto -1/x$ in the
Boltzmann regime ($x \gg x^*\approx 10$).
On the other hand, in the quantum regime $x \ll x^*$ (Fig. \ref{f-vs-x}a), $f_3$ shows a significant dependence
on  $\mu_0$, signalizing that the nonlinear dc transport properties
are no more governed uniquely by the semiclassical parameter $x$ but that
the detailed value of $\mu_0$ (or conversely, of $U$) starts
playing a relevant role.
Some striking things are worth being pointed out: ($i$)
counterintuitively, the third-order contribution to the dc transport
appears to be magnified approaching the clean limit $U \rightarrow 0$;
($ii$) there are two isosbestic points (i.e. $x$-points where $f_3$ does not depend on $U$)
coinciding in a very good approximation with the zeroes of the $f_3$ function;
($iii$) whereas the phenomenological constant-$\Gamma$ modelling predicts a
a well-determined positive sign of the nonlinear dc correction in the quantum regime (implying an increase
of the total conductivity), the sign of the nonlinear terms of the full conserving many-body theory
in the quantum regime is not univocally determined, presenting a positive region in the crossover range and a negative sign in the extreme quantum limit.

\begin{figure}[t]
\includegraphics[width=90mm]{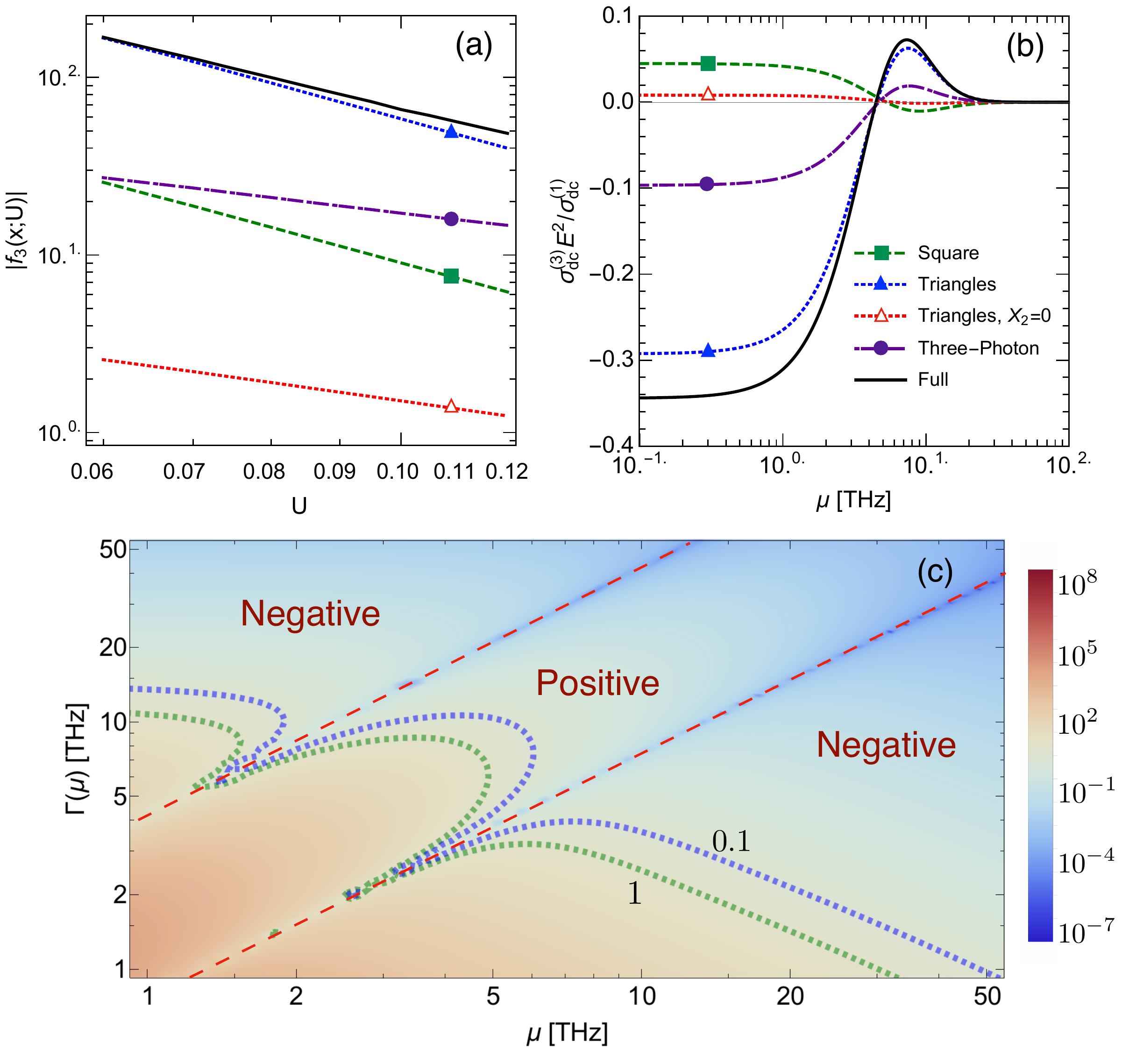}
\caption{
{\bf Nonlinear conductivity of graphene at zero temperature.} 
(a) Log-log scale plot for the absolute value of the characteristic third-order transport function $f_3(\mu,U)$ versus $U$ at $x =0.03$.
Different lines correspond to the individual contribution of diagrams in Fig.~\ref{fig:chi3}a as mentioned the shared plot-legend in panel (b). 
(b) $f_3(\mu,U)$ as a function of the chemical potential for $U=0.11$. Similar to the panel (a), different curves correspond to different diagram's contribution as mentioned the plot-legend. (c) Colormap plot for $g=|\sigma^{(3)}_{\rm dc} E^2/\sigma^{(1)}_{\rm dc}|$ factor with $ E=1{\rm mV/nm}$ versus chemical potential $\mu$ and relaxation rate $\Gamma(\mu)$ in the full conserving models. The sign of $\sigma^{(3)}_{\rm dc}$ is written on the plot where two sign-switch borders are highlighted by dashed red lines. Green and blue dashed lines stand for the contour lines with $g=1$ and $g=0.1$, respectively.  Similar colormap plot for the constant-$\Gamma$ model is given in the SM \cite{SM}.
}
\label{f-panels}
\end{figure}

We can rationalize points ($i$)-($ii$) by assuming that in the quantum regime
the nonlinear characteristic dc transport function $f_3(x,U)$ can be factorized as \cite{SM}:
\begin{eqnarray}
f_3(x;U) 
&\approx&
\frac{C(x)}{U^\gamma},
\label{eqscaling}
\end{eqnarray} 
(with $\gamma >0$), where the strength $U$ of the interaction rules
governs the {\em intensity} of the third-order dc transport,
while the semiclassical parameter $x$ seems to dictate the {\em sign}
of the third-order correction.
To assess the meaningfulness of such description we plot
in Fig. \ref{f-panels}a in a log-log scale the absolute value of the function $f_3(x,U)$
versus $U$ for a representative case $x=0.03$ in the quantum regime.
We find a perfect agreement with a scaling behavior $f_3(x,U)\propto 1/U^\gamma$
 with $\gamma$ slightly smaller than 2
($\gamma = 1.82$) signalizing that in the clean limit the third-order dc transport is expected
to be dominant with respect to the linear one.
As detailed in \cite{SM}, a similar analysis is valid in the whole quantum regime.
 
In order to gain a full understanding of these novel features,
we analyze separately in Fig. \ref{f-panels}a,b the relevance of each family of diagrams
contributing to the total third-order conductivity
as depicted in Fig. \ref{fig:chi3}a.
We can thus realize that the contribution of the conventional ``square'' diagram
(which is the only one present in the non-interacting case
and for the phenomenological damping model), 
is essentially marginal, as well as
the contribution of the last ``bubble''
associated with the renormalized three-photon vertex.
The dominant role is instead played by the ``triangle'' diagrams
containing the renormalized {\em two}-photon vertex.
A quantitative analysis shows that each family of diagrams
obeys Eq. (\ref{eqscaling}) with an approximately integer exponent
(i.e. $\gamma_{\mdblksquare}\approx 2$, 
$\gamma_{\blacktriangle}\approx 2$, and $\gamma_{\sbullet[1.2]}\approx 1$ for the square, triangle and the bubble diagrams).
The dominance of the triangle diagrams results thus in an exponent very close
to 2 ($\gamma \approx \gamma_{\blacktriangle}$).
The self-consistent BS renormalization
of the TPV (Fig. \ref{fig:chi3}c)
is a crucial ingredient in such novel scenario.
This can be assessed in Fig. \ref{f-panels}a,b
where one can see that,
once neglected the BS renormalization,
the contribution of the triangle diagrams results to be of the same order (even smaller)
of that of the conventional square diagram.
The dominant role of the TPV renormalization appears
even more evident by investigating the scaling of the
characteristic third-order dc
transport function $f_3(x,U)$ versus $U$.
As depicted in Fig. \ref{f-panels}a, 
once replaced the BS renormalized TPV (Fig. \ref{fig:chi3}c)
with the ``bare'' one (Fig. \ref{fig:chi3}e), 
the third-order dc conductivity scales as $1/U$ ($\gamma_{\triangle}\approx 1$),
with an additional sign change change, as shown in  Fig. \ref{f-panels}b.
This means that the BS renormalization of the TPV
gives rise in the quantum regime to an additional dependence $\sim 1/U$ that diverges in the clean limit. The $n$-photon vertex matrix structure, which reads $\hat \Lambda_{n} = (-ev \hat \sigma_y)^n \Lambda_n$, plays a crucial role in the relevance of the BS renormalization effect. 
The impressive effect is peculiar of the TPV renormalization $\Lambda_2=\Lambda_2^{(0)}/[1-UX_2]$ and 
does not appear in the BS renormalization of
the one- three-photon vertex ($\Lambda_n=\Lambda_n^{(0)}/[1-UX_n]$, with $n=1,3$) \cite{SM}.
This different impact can be traced down to the different structure in the Pauli space.
As detailed in \cite{SM},
we get indeed
$X_1=X_3 \propto {\rm Tr}[\hat \sigma_y \hat G \hat \sigma_y \hat G ]$,
$X_2 \propto {\rm Tr}[\hat G \hat G ]$.
In the quantum regime, one can thus show that  in the dc limit $UX_1\approx UX_3 \propto U$,
whereas $UX_2 \approx 1+{\cal O}(U)$, so that
resulting in an effective divergence of $\Lambda_2/\Lambda^{(0)}_2$ in the dc limit
at zero temperature and in the clean limit ($U\to0$). 

The impact of the two-photon renormalization can be understood in more details by investigating
the two-photon renormalization factor, \cite{SM}
\begin{equation}
\label{eq:Q2}
Q_2(z_1,z_2) =\frac{1}{1-U X_2(z_1,z_2)} = \frac{S(z_1)-S(z_2)}{z_1-z_2}~,
\end{equation}
where  $X_2(z_1,z_2)\propto\sum_{\bf k}{\rm Tr}[\hat G({\bf k},z_1) \hat G({\bf k},z_2)]$ and
where $z_1$ and $z_2$ are the electronic frequencies in the complex plane. 
Particularly enlightening is the analysis of the retarded-retarded (RR) channel.
In the dc limit ($\omega\to 0$)  $Q^{\rm RR}_2(\epsilon,\epsilon+\omega)$  is simply given by 
\begin{align}
\label{eq:RR_vertex}
\lim_{\omega \to 0}
Q^{\rm RR}_2(\epsilon,\epsilon+\omega)
= \frac{dS(\epsilon)}{d\epsilon} = \frac{S(\epsilon)}{2US(\epsilon)+\mu_0+\epsilon},
\end{align}
where $\omega$ is the photon energy and
$\epsilon$ is the electronic energy from the Fermi surface.
For low-energy excitations $\epsilon=0$ we have thus
$Q^{\rm RR}_2=S(0)/[2US(0)+\mu_0]$.
The Boltzmann regime is achieved as $\mu_0 \gg 2US(0)$. In the clean limit $U\to 0$ we get thus
$S(0)=\mu_0$ and $Q^{\rm RR}_2=1$.
The quantum regime is on the other hand characterized by $\mu_0 \ll 2US(0)$,
and we get $Q^{\rm RR}_2=1/2U$, leading thus to a huge {\em enhancement} (divergence) in the clean limit.
A similar behavior $Q_2\propto 1/U$ appears also in the retarded-advance (RA) channel,
although is a more delicate way.
In the dc limit, we find indeed the leading term in $Q^{\rm RA}_2(\epsilon,\epsilon+\omega) \approx -i 2\Gamma(\epsilon)/\omega$
which shows a divergence as a function of the photon energy $\omega \to 0$.
Once plugged this behavior in the response function associated with the ``triangle'' diagrams containing
the renormalized BS two-phonon vertex,
$\rchi^{\rm ren.}_{\rm triangles}(\omega)\sim \Gamma(\epsilon) \rchi^{\rm unren.}_{\rm triangles}(\omega)/\omega$
the divergence is $\omega$ implies that $\rchi^{\rm unren.}_{\rm triangles}(\omega)$
must be expanded to a higher order in $\omega$, involving the {\em second} derivative
$S''(0) = - 1/[4U^2 \Gamma(0)]$ and higher orders.
As a net result, the divergence $1/\omega$ in $Q^{\rm RA}_2(\epsilon,\epsilon+\omega)$ is reflected
in a consequent divergence $-1/U$ in the response function, similarly as for the RR channel, but with a {\em negative} sign.
The balance between RR and RA terms determines the change of sign
of the third-order dc conductivity as a function of $x$ n the quantum regime.
We must stress that the huge enhancement of the third-order dc transport is governed
by the dominant role of the two-phonon vertex renormalization.
Since $\Lambda_2^{(0)}$ scales as $U$ and $1/[1-UX_2]$ scales as $1/U$,
such enhancement can be regarded as TPV self-generation ($\Lambda_2\neq 0$)
which survives in the weak-coupling (clean) limit $U \to 0$.

The net result on the dc transport is summarized in Fig. \ref{f-panels}c
where we plot  the sign and the magnitude of the third-order conductivity
for a given electric field $E=1$ mV/nm normalized to the linear order conductivity, 
$g=|\sigma^{(3)}_{\rm dc} E^2/\sigma^{(1)}_{\rm dc}|$,
in the physical space of the effective chemical potential $\mu$
and scattering rate $\Gamma(\mu)$, as they can be obtained directly
in an experimental way.
The Boltzmann regime corresponds thus to the right-lower corner
whereas the extreme quantum regime ($x \to 0$) is recovered in the left-upper corner.
As noticed before, at odds with the predictions of the phenomenological model,
we find that the third-order conductivity $\sigma^{(3)}_{\rm dc}$ is negative not only in the Boltzmann regime,
but also in the {\em quantum} regime.
Note that the zeroes of the third-order dc conductivity, see $x_1$ and $x_2$ in Fig.~\ref{f-vs-x}, 
appear in this plot as straight dashed lines. This is a consequence of the
factorizable expression for the characteristic third-order
dc transport function as shown in Eq. (\ref{eqscaling}).
We mark with tiny dotted in this plot the regions where
the third-order terms start to be relevant $g \approx 0.1$
and where they become of the same order than the linear dc term $g \approx 1$ \cite{noteg1}.

Note that in the Boltzmann regime, by increasing $\mu \to \infty$, one should correspondingly need
infinitesimally small $\Gamma \to 0$ in order to detect third-order corrections, while
in the quantum regime $\mu \to 0$ third-order corrections appear to be relevant
up to large values of $\Gamma$ dictated only by the applied electric field
(and by the ultra-violet cut-off of the Dirac model).
An alternative and maybe more direct way to assess the relevance of the nonlinear conductivity is to evaluate
in graphene \cite{notegraphene} as a paradigmatic 2D Dirac material
the critical electric field $E_{\rm max}$ above which third-order corrections
to the dc transport become of the same order of the linear term,
$|\sigma^{(3)}_{\rm dc}| E^2_{\rm max} \sim \sigma^{(1)}_{\rm dc}$.
In the phenomenological constant-$\Gamma$ model we obtain $E_{\rm max} = \alpha E_0$ with
$\alpha \approx1.47\times \mu \Gamma/t^2_0$
in the Boltzmann regime and $\alpha \approx 1.42\times \Gamma^2/t^2_0$
in the quantum limit. 
These values can be compared with the estimates for the full conserving theory
that gives $\alpha \approx 0.32 \times \mu \Gamma(\mu)/t^2_0$
in the Boltzmann regime and $\alpha \approx 1.14 U \times \Gamma(0)^2/t^2_0$
in the quantum limit with $U=1/\ln[W^2/\Gamma^2(0)]$.
In Ref. \cite{Horng_prb_2011} a roughly constant value $\Gamma\approx 15$ meV
was estimated in the wide range $\mu\sim 0-200$ meV.
With these values the phenomenological model would estimate a critical field $E_{\rm max}\approx 10.8$ mV/nm
for $\mu \approx 200$ meV and $E_{\rm max}\approx 74.7$ $\mu$V/nm for $\mu=0$
with a quantum-Boltzmann crossover at $\mu^*\approx 150$ meV,
whereas the full conserving theory predicts $E_{\rm max}\approx 2.35$ mV/nm
for $\mu \approx 200$ meV and $E_{\rm max}\approx 0.051$ mV/nm for $\mu=0$,
in a more observable range.
 
In conclusion, in this Letter, we have presented a fully conserving theory of nonlinear transport response
in 2D Dirac materials. Our results show that the previous analyses in literature, based on phenomenological scattering models,
can be qualitatively (but not quantitatively) reliable in the Boltzmann regime but they completely fail in the quantum regime.
We have shown that, in a wide region of the phase diagram, close to the neutral point,
the nonlinear dc transport response is dominated by novel physical processes where
the two-photon vertex, absent in the bare Dirac Hamiltonian, plays a relevant role.
It should be furthermore stressed that our results, focused on the dc limit, implies that
 the current knowledge about the nonlinear optical response in the terahertz regime should be deeply revised.
Our work opens new scenarios for a deep understanding of the electromagnetic response
of Dirac systems, whose relevance ranges from condensed matter to high-energy physics.

{\it Acknowledgements--.} H.R. acknowledges the support from the Swedish Research Council (VR 2018-04252).
\bibliography{bibliography}
\widetext
\vspace{20cm}

\setcounter{equation}{0}
\setcounter{figure}{0}
\setcounter{table}{0}
\makeatletter

 \renewcommand{\theequation}{S\arabic{equation}}
\renewcommand{\thesection}{S\arabic{section}}
\renewcommand{\thefigure}{S\arabic{figure}}
  
\begin{center}
\textbf{\large Supplemental Materials: \\ ``Dominant Role of Two-Photon Vertex in Nonlinear Response of Dirac Materials''\\}
\vspace{0.5cm}
Habib Rostami,$^{1}$
Emmanuele Cappelluti$^{2}$\\[4pt]
$^{1}${\small\it Nordita, KTH Royal Institute of Technology and Stockholm University, Roslagstullsbacken 23, SE-106 91 Stockholm, Sweden}
$^{2}${\small\it Istituto di Struttura della Materia, CNR, 34149 Trieste, Italy}
\end{center}
\date{\today}
\tableofcontents
\section{Ultra-violet cut-off and dimensional regularization}\label{sec:dim_reg}
The introduction of a high-energy (ultra-violet) cut-off is an unavoidable requirement
of Dirac models. There is however a relative large degree of freedom in
the way how to introduce it, and particular care is needed in order to avoid
spurious results and to preserve physical consistencies, like Ward's identities, and gauge invariance.
Dimensional regularization has proven to be a formidable tool
to ensure that physical correctness is preserved \cite{Leibbrandt_rmp_1975,Peskin}.
Here we show how such approach provides a consistent framework for evaluating self-energy and susceptibilities. In our work we employ dimensional regularization that endures gauge invariance.
As a benchmark example, and for the sake of simplicity,
we consider the evaluation of the disorder self-energy
which displays a primary diverging integral. 
 
 We consider scattering on local impurity centers
with density $n_{\rm imp}$ and potential $V_{\rm imp}({\bf r})=\sum_iV_i\delta({\bf r}-{\bf R}_i)$
where ${\bf R}_i$ are the coordinates of the lattice sites.
We assume standard Born impurity correlations as $\langle V_{\rm imp}({\bf r}) \rangle=0$ and 
the effective scattering potential reads 
\begin{align}
V(1,2) = \langle V_{\rm imp}({\bf r}_1) V_{\rm imp}({\bf r}_2)  \rangle_{\rm imp}   =\gamma_{\rm imp} \delta({\bf  r}_1-{\bf r}_2)~.  
\end{align}
Note that the average $\langle \dots \rangle_{\rm imp}$ is meant over all the impurity configurations and $\gamma_{\rm imp} = n_{\rm imp} V^2_{\rm imp}$ in which $n_{\rm imp} = N_{\rm imp}/N_{\rm cell}$ (number of impurity centers per number of unit-cells) stands for the density of scattering centers and  $V_{\rm imp}$ is the average strength of the scattering potential energy. The lowest-order self-consistent Born self-energy reads
\begin{align}
\hat \Sigma(ik_n) = \gamma_{\rm imp} \sum_{\bf k} \hat G({\bf k},ik_n)  = 
\gamma_{\rm imp} {\cal S}  \int \frac{d^2k}{(2\pi)^2}\hat G({\bf k},ik_n)~. 
\end{align}
Note that ${\cal S}= N_{\rm cell} S_{\rm cell}$ is the system area with $S_{\rm cell}$ being the unit-cell area. Due to the isotropic impurity scattering the self-energy spinor structure is trivial as  $\hat \Sigma(ik_n)= \Sigma(ik_n)\hat I$ and therefore the Green's function can be explicitly written as follows 
\begin{figure}[b]
\includegraphics[width=180mm]{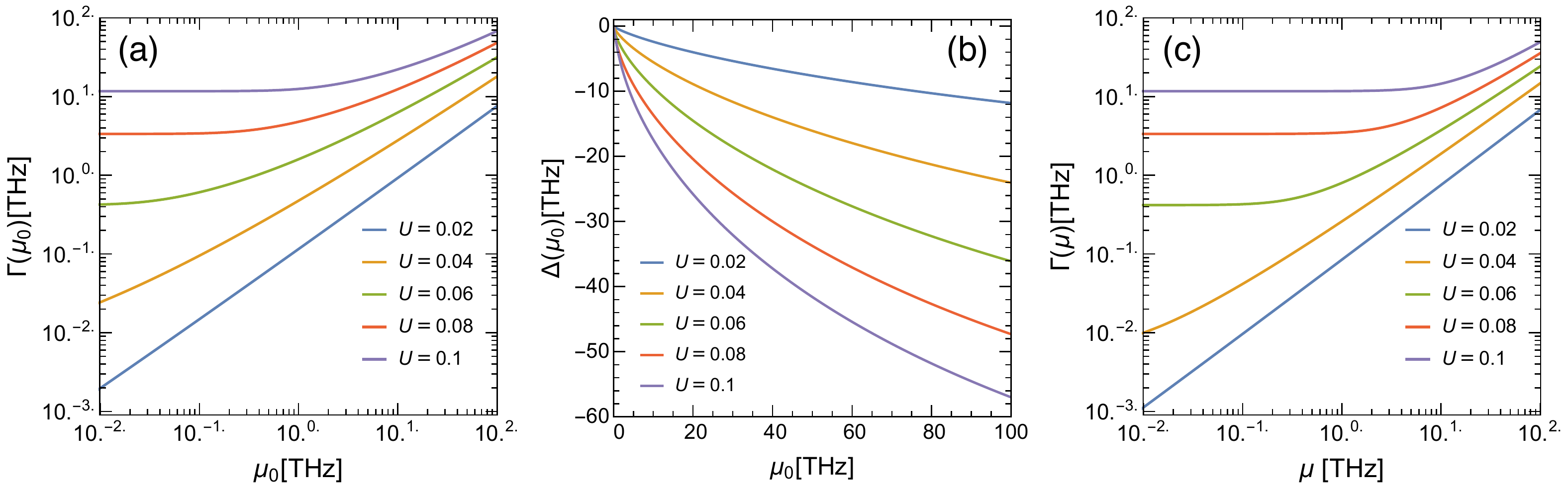}
\caption{{\bf Numerical result for the imaginary and real parts of the self-energy.} (a) The imaginary part of the self-energy at the Fermi surface $\Gamma = - {\rm Im}[\Sigma(\omega=0)]$ is shown versus bare chemical potential $\mu_0$. 
(b) The imaginary part of the self-energy at the Fermi surface $\Delta =  {\rm Re}[\Sigma(\omega=0)]$ is shown versus bare chemical potential $\mu_0$. (c)  $\Gamma$ is shown versus renormalized chemical potential $\mu =\mu_0 -\Delta(\mu)$. Different curves correspond to different values of $U$.}
\label{fig:selfenergfy}
\end{figure}
\begin{align}
\hat G({\bf k},ik_n) = \frac{[z+\mu_0-\Sigma(z)]\hat I+ \hbar v 
\hat{\bm \sigma} \cdot{\bf k}}{[z+\mu_0-\Sigma(z)]^2-(\hbar v k)^2} 
\end{align}
where $z=ik_n$. Accordingly, we find 
\begin{align}
\Sigma(z) = \frac{\gamma_{\rm imp} N_{\rm cell} S_{\rm cell}}{(\hbar v)^2}  \int \frac{d^2 \ell}{(2\pi)^2} \frac{S(z) }{S(z)^2-\ell^2}
\end{align}
where $S(z) = z+\mu_0-\Sigma(z)$. In arbitrary D dimensions, we have 
\begin{align}
\Sigma(z) = \frac{\gamma_{\rm imp} N_{\rm cell} S_{\rm cell}}{(\hbar v)^D}  \int \frac{d^D \ell}{(2\pi)^D} \frac{S(z) }{S(z)^2-\ell^2}~.
\end{align}
Note that the above integral in D dimensions can be solved in terms of Euler's Gamma-function, $\Gamma_E(z)$, by utilizing the following identity \cite{Peskin}
\begin{align}
\int \frac{d^D \ell}{(2\pi)^D} \frac{1}{(\ell^2+\Delta)^n} = 
\frac{1}{(4\pi)^{\frac{D}{2}}} \frac{\Gamma_E(n-\frac{D}{2})}{\Gamma_E(n)} 
\left(\frac{1}{\Delta}\right)^{n-\frac{D}{2}}~.
\end{align}
Therefore, we find 
\begin{align}
\Sigma(z) = -U S(z)
\frac{1}{(4\pi)^{\frac{D}{2}-1}(\hbar v)^{D-2}} 
\Gamma_E\Big(1-\frac{D}{2}\Big) \left(-S(z)^2\right)^{\frac{D}{2}-1}
\end{align}
in which  \cite{notegraphene}
\begin{align}
 U = \frac{ \gamma_{\rm imp} N_{\rm cell} S_{\rm cell}}{4\pi (\hbar v)^2}
 = \frac{N_{\rm imp} }{4\pi\sqrt{3}} \left(\frac{V_{\rm imp}}{t_0}\right)^2~. 
\end{align}
Now we set $D=d-\epsilon$ where $d=2$ is the physical dimension and $\epsilon\to 0$. 
Note that $\Gamma_E(\epsilon/2) \approx  2/\epsilon$ for  $\epsilon\to 0$ and
\begin{align}
\lim_{\epsilon \to 0} \frac{(X^2)^{-\epsilon/2}}{\epsilon/2} = 
\lim_{\epsilon \to 0} \frac{2}{\epsilon} -\ln[X^2] = \ln[W^2] -\ln[X^2]= \ln \left[\frac{W^2}{X^2}\right]~.
\end{align}
Note that we use the prescription $\lim_{\epsilon \to 0} \frac{1}{\epsilon}  \equiv \ln[W]$ where $W$ is the ultra-violet energy cut-off. Eventually, we obtain the following self-consistent formula for the self-energy 
\begin{align}\label{eq:self}
\Sigma (z) = - U S(z)\ln\left[- \frac{W^2}{S(z)^2} \right]~.
\end{align}
After solving the above relation self-consistently, the real, $\Delta =  {\rm Re}[\Sigma(\omega=0)]$,  and imaginary, $\Gamma = - {\rm Im}[\Sigma(\omega=0)]$, parts of the self-energy at the Fermi surface are depicted in Fig.~\ref{fig:selfenergfy}. For the undoped regime and at the Dirac point, we have $\Sigma(0) = -i\Gamma(0)$ where 
\begin{align}
\Gamma(0) = W \exp\left\{-\frac{1}{2U}\right\}~.
\end{align}
The above procedure of dimensional regularization is employed in similar way
in the evaluation of other momentum integrals in this study.
 \section{Baym-Kadanoff derivations} \label{sec:KB}
Within the lowest-order self-consistent Born approximation,  the self-energy correction induced by elastic impurity scattering reads  
\begin{align}\label{eq:scba}
  \hat \Sigma(1,2) = V(1,2)   \hat G(1,2)~.
\end{align}
We use a shorthand notation $1\equiv({\bf r}_1,t_1)$ for the space-time coordinate.
Using Dyson recursive relation, the full field-dependent and interacting Green's function is given in terms a 
 field-dependent self-energy, $\hat \Sigma$, and a bare Green's function, $\hat G_0$, as follows 
\begin{align}
   \hat G(1,1';{\bf A}) =   \hat G_0(1,1';{\bf A})   + \int_{\bar2,\bar3}  \hat G_0(1,\bar 2;{\bf A})   \hat \Sigma(\bar 2,\bar 3;{\bf A})    \hat G(\bar3,1';{\bf A}) 
\end{align}
or equivalently we have
\begin{align}\label{eq:dyson}
  \hat G^{-1}(1,1';{\bf A}) =  \hat G^{-1}_0(1,1';{\bf A})-  \hat\Sigma(1,1';{\bf A})~.
\end{align}
The inverse of bare Green's function reads 
\begin{align}
  \hat G^{-1}_0(1,1';{\bf A}) &=  [i\partial_{t_1}- v   \hat{\bm \sigma}\cdot(-i\hbar {\bm \nabla}_1 + e{\bf A}(1)) ] \delta(1-1')~.
\end{align} 
We assume an external gauge field along y axis, ${\bf A}(1) = A(1) \hat{\bf y}$. 
The one-photon current vertex is given in terms of variational derivative of bare Green's function versus the gauge field:
\begin{align}
 \hat \Lambda^{(0)}_{1}(1,1';1'') = \frac{\delta   \hat G^{-1}_0(1,1')}{\delta A(1'')}\Big|_{\bf A\to 0}
 = -e v \hat\sigma_y \delta(1-1'') \delta(1-1')~.
\end{align}
The thermodynamic physical current, ${\bf J}(1;{\bf A})= J(1;{\bf A}) \hat {\bf y}$, in Dirac systems reads
\begin{align}\label{eq:physical_current}
J(1;{\bf A}) = 
-i \int_{1',1''} {\rm tr} \left[  \hat \Lambda^{(0)}_{1} (1,1';1'')    \hat  G(1,1'^+;{\bf A})  \right]~.
\end{align}
Note that $1'^+ \equiv ({\bf r}_{1'}, t_{1'}=t_{1}+0^+)$ and ``${\rm tr}$'' stands for the ``trace'' operation over all spinor indexes i.e. ${\rm tr}[\hat A\hat B]=\sum_{ss'}[A_{ss'}B_{s's}]$.  Note the field operator $\hat \psi_{\cal H}({\bf r},t) $ in the Heisenberg picture of $\hat \psi({\bf r}) $ in the basis of full Hamiltonian ${\cal H}$ which contains kinetics, light-matter and many-body interaction terms.  The Baym-Kadanoff (or contour) Green's function \cite{Baym_Kadanoff_prb_1961,Baym_Kadanoff_book} follows 
\begin{align}
 \hat G(1,1';{\bf A}) &=-i \langle {\cal T}[ \hat \psi_{\cal H}(1)\hat \psi^\dagger_{\cal H}(1') ] \rangle~. 
\end{align}
where $\langle \dots \rangle$ stands for the thermodynamical average and $ {\cal T}$ is for the time-ordering operation. 
\subsection{Conserving linear response theory in Dirac systems}
We define the linear susceptibility as follows 
\begin{align}
\rchi^{(1)}  (1,2) &= \frac{\delta J(1;{\bf A}) }{\delta A(2)} \Big|_{{\bf A} \to 0}~, 
\label{eq:chi1_def} 
\end{align}
Using Eq.~(\ref{eq:physical_current}), the linear susceptibility reads
\begin{align}
&\rchi^{(1)}(1,2) 
= 
-i\int_{1',1''}~{\rm tr}\left[\hat \Lambda^{(0)}_1(1,1';1'') 
\frac{\delta   \hat G(1,1'^+;{\bf A}) }{\delta A(2)} \right]_{{\bf A}\to 0} 
\end{align}
Using $\hat G \hat G^{-1}=1$, we have
\begin{align}\label{eq:dG}
\frac{\delta  \hat G(1,1';{\bf A})}{\delta A(2)} = - \int_{\bar 2,\bar 3} 
  \hat G(1,\bar2;{\bf A})\frac{\delta \hat G^{-1}(\bar2,\bar3;{\bf A})}{\delta A(2)} \hat  G(\bar3,1';{\bf A})
\end{align}
Using Eq.~(\ref{eq:dyson}), we find
\begin{align}
\frac{\delta  \hat G^{-1}(1,1';{\bf A})}{\delta A(2)} =  \frac{\delta \hat  G^{-1}_0(1,1';{\bf A})}{\delta A(2)} - \frac{\delta   \Sigma(1,1';{\bf A})}{\delta A(2)} 
\end{align}
Since the self-energy depends on the external potential only through its dependance on the Green's function, we can write down 
\begin{align}
\frac{\delta  \hat \Sigma(1,1';{\bf A})}{\delta  A(2)}
=\int_{\bar 3,\bar4}~ 
\hat \Xi (1,1' ; \bar3,\bar4;{\bf A})
\frac{\delta  \hat G(\bar3,\bar4;{\bf A})}{\delta A(2)}
\end{align}
We define {\it Bethe-Salpeter kernel} as bellow 
\begin{align}
\hat \Xi (1,2; 3,4;{\bf A})=  \frac{\delta \hat \Sigma(1,2;{\bf A})}{\delta  \hat G(3,4;{\bf A})}
\end{align}
Note that $ \hat G (1,1';{\bf A})|_{\bf A \to 0} =  \hat G (1,1')$ and 
$ \hat \Xi (1,2;3,4;{\bf A})|_{\bf A \to 0} =  \hat \Xi (1,2;3,4)$. For our self-energy model within theself-consistent Born approximation given in Eq.~(\ref{eq:scba}), we have 
\begin{align}
\hat \Xi(1,2; 3,4)= V(1,2) \delta(1-3) \delta(2-4) 
\end{align}
Therefore, using  Eq.~(\ref{eq:dG}) the self-consistent Bethe-Salpater relation for the one-photon vertex function  follows
\begin{align}
\hat \Lambda_1(1,1';2)
= \hat\Lambda^{(0)}_1(1,1';2) +  V(1,1')  \int_{\bar5,\bar6} \hat G(1,\bar5)\hat \Lambda_1 (\bar5,\bar6;2) \hat G(\bar6,1')
\end{align}
Note that the dressed one-photon vertex function are defined as  
\begin{align}\label{eq:ver1}
\hat\Lambda_1(1,1';2)=\frac{\delta \hat G^{-1}(1,1';{\bf A})}{\delta A(2)}\Big|_{\bf A\to 0}
\end{align}
For the sake of simplicity, we extend the definition of space-time parameter to include also spinor indexes as $\bar1=({\bf r}_{\bar1},t_{\bar1},s_{\bar1})$ and, for instance, we can drop ``$\hat{~} $'' symbol in $\hat \Lambda_1 \hat G$ instead use $  \Lambda_1  G$ since the spinor multiplication is taken into account when we replace $\int d\bar1\sum_{s_{\bar1}} \to \int d\bar1$. Moreover we use shorthand notation for  $C =\int d\bar1 A(\dots,\bar1) B(\bar1,\dots)$ as $C=AB$. We use this compact notation from now on 
\begin{align}
\rchi^{(1)}_{\alpha\beta}(1,2) &=i{\rm Tr}\left[ \Lambda^{(0)}_1(1)  G\Lambda_1(2)  G \right]
\label{eq:chi1}
\\
 \Lambda_1(2) &= \Lambda^{(0)}_1(2) + V G  \Lambda_1(2) G
\label{eq:sc_ver1}
\end{align}
Note that ${\rm Tr}[\dots]$ in the above formula stands for the sum over un-contracted spinor index. In the compact notation $1$ and $2$ are space-time symbols. 
\subsection{Conserving third order response theory in Dirac systems}
Third-order response function is given by 
\begin{align}
\rchi^{(3)} (1,2,3,4)\equiv \frac{1}{3!}\frac{\delta^3 J(1;{\bf A})}{\delta A(2) \delta A(3)\delta A(4)}\Big|_{{\bf A}\to 0} 
=-\frac{i}{3!}{\rm Tr}\left[ \Lambda^{(0)}_1(1) \frac{\delta^3 G(1,1^+;{\bf A})}{\delta A(2) \delta A(3)\delta A(4)}  \right]_{{\bf A}\to 0}
\end{align}
Using $  G G^{-1}=1$  and Eq.~(\ref{eq:dG}), we evaluate the third derivative of the Green's function versus vector potential component,  
 \begin{align}
&\frac{\delta^3   G}{\delta A(2)\delta A(3) \delta A(4)} =
-\frac{\delta^2   G}{\delta A(2)\delta A(3) } \frac{\delta   G^{-1} }{ \delta A(4)}  G
-\frac{\delta^2   G}{\delta A(3)\delta A(4)} \frac{\delta   G^{-1}}{\delta A(2)}  G
+  G \frac{\delta   G^{-1}}{\delta A(3)}   G\frac{\delta^2   G^{-1}}{\delta A(2)\delta A(4)}  G
\nonumber\\&- \frac{\delta^2   G}{\delta A(2)\delta A(4)} \frac{\delta   G^{-1}}{\delta A(3)}  G
+  G\frac{\delta   G^{-1}}{\delta A(2)}   G\frac{\delta^2   G^{-1}}{\delta A(3)\delta A(4)}  G
+  G\frac{\delta   G^{-1}}{\delta A(4)}   G\frac{\delta^2   G^{-1}}{\delta A(2) \delta A(3)}  G
-  G \frac{\delta^3   G^{-1}}{\delta A(2) \delta A(3)\delta A(4)}   G~.
\end{align}
Using  Eq.~(\ref{eq:ver1}), we have
 \begin{align}\label{eq:d3GA3}
&\frac{\delta^3   G}{\delta A(2)\delta A(3) \delta A(4)} =
-\frac{\delta^2   G}{\delta A(2)\delta A(3) }  \Lambda_1(4) G
-\frac{\delta^2   G}{\delta A(3)\delta A(4)}  \Lambda_1(2) G
+  G  \Lambda_{1}(3)    G\frac{\delta^2   G^{-1}}{\delta A(2)\delta A(4)}  G
\nonumber\\&- \frac{\delta^2   G}{\delta A(2)\delta A(4)}  \Lambda_1(3)G
+  G \Lambda_{1}(2)     G\frac{\delta^2   G^{-1}}{\delta A(3)\delta A(4)}  G
+  G  \Lambda_{1}(4)      G\frac{\delta^2   G^{-1}}{\delta A(2) \delta A(3)}  G
-  G \frac{\delta^3   G^{-1}}{\delta A(2) \delta A(3)\delta A(4)}   G~.
\end{align}
Once more we use $  G G^{-1}=1$ and perform second derivative of the Green's function which leads 
\begin{align}
\frac{\delta^2   G}{\delta A(2)\delta A(3)} =
-\frac{\delta   G}{\delta A(3)} \frac{\delta   G^{-1}}{\delta A(2)}  G
- \frac{\delta   G}{\delta A(2)} \frac{\delta   G^{-1}}{\delta A(3)}   G
-  G \frac{\delta^2   G^{-1}}{\delta A(2) \delta A(3)}   G~.
\end{align}
The two-photon vertex function is defined as follows   
\begin{align}\label{eq:ver2}
 \Lambda_2(2,3) =   \frac{\delta^2   G^{-1}}{\delta A(2) \delta A(3)}~.
\end{align}
Using Eq.~(\ref{eq:dG}), Eq.~(\ref{eq:ver1}) and Eq.~(\ref{eq:ver2}), we find   
 \begin{align}\label{eq:d2G}
 \frac{\delta^2   G}{\delta A(2)\delta A(3)} =-  G  \Lambda_2(2,3)   G
+  \sum_{{\cal P}(2;3)}   G  \Lambda_1(2)   G  \Lambda_1(3)   G
\end{align}
where ${\cal P}(2;3)$ stands for the permutation between $2\leftrightarrow 3$. 
The three-photon vertex function reads 
\begin{align}\label{eq:ver3}
 \Lambda_3(2,3,4) =  \frac{1}{2} \frac{\delta^2   G^{-1}}{\delta A(2) \delta A(3) \delta A(4)}~.
\end{align}
By plugging Eq.~(\ref{eq:ver2}) , Eq.~(\ref{eq:d2G}), and Eq.~(\ref{eq:ver3}) in Eq.~(\ref{eq:d3GA3}), we obtain
  \begin{align}\label{eq:d3G}
\frac{\delta^3   G}{\delta A(2)\delta A(3) \delta A(4)} 
&=
-2  G   \Lambda_3(2,3,4)   G
-\sum_{{\cal P}(2;3;4)}   G  \Lambda_1(2)   G  \Lambda_1(3)  G  \Lambda_1(4)  G
+\frac{1}{2}\sum_{{\cal P}(2;3;4)}  G   \Lambda_2(2,3)   G  \Lambda_1(4)  G
\nonumber\\&
+\frac{1}{2}\sum_{{\cal P}(2;3;4)}  G    \Lambda_1(4)  G   \Lambda_2(2,3)  G~.
\end{align}
where ${\cal P}(2;3;4)$ stands for all six permutations among 2,3, and 4 space-time coordinates.
Therefore, the third-order response function reads
\begin{align}\label{eq:chi3}
\rchi^{(3)}(1,2,3,4)
&=2\frac{i}{3!}{\rm Tr}\left[ \Lambda^{(0)}_1(1)   G   \Lambda_3(2,3,4)   G  \right]
+\frac{i}{3!} \sum_{{\cal P}(2;3;4)} {\rm Tr}\left[ \Lambda^{(0)}_1(1) G \Lambda_1(2) G  \Lambda_1(3) G \Lambda_1(4) G \right]
\nonumber\\ &
-\frac{1}{2}\frac{i}{3!} \sum_{{\cal P}(2;3;4)}{\rm Tr}\left[ \Lambda^{(0)}_1(1)   G   \Lambda_2(2,3)   G  \Lambda_1(4)  G  \right]
-\frac{1}{2}\frac{i}{3!} \sum_{{\cal P}(2;3;4)}{\rm Tr}\left[ \Lambda^{(0)}_1(1)  G    \Lambda_1(4)  G   \Lambda_2(2,3)  G  \right]~.
\end{align}
\subsubsection{Interaction induced two-photon vertex}
The two-photon vertex function is defined as follows   
\begin{align} 
   \Lambda_2(2,3) =   \frac{\delta^2   G^{-1}}{\delta A(2) \delta A(3)} 
=   \frac{\delta^2   G^{-1}_0}{\delta A(2) \delta A(3)} -  \frac{\delta^2   \Sigma}{\delta A(2) \delta A(3)}~. 
\end{align}
The second derivative of the bare Green's function vanishes in Dirac systems which implies the absence of bare two-photon vertex. However, an interaction induced two-photon vertex function is obtained owing to the field-dependent self-energy. Since the self-energy depends on the external field only through the dependence on the Green's function, we have 
\begin{align}
\Lambda_2(2,3)
=- \frac{\delta^2   \Sigma}{\delta A(2)\delta A(3)} 
= - V \frac{\delta^2   G}{\delta A(2)\delta A(3)}~.
\end{align}
Using Eq.~(\ref{eq:d2G}), we obtain
\begin{align}\label{eq:sc_ver2}
 \Lambda_2(2,3) =  \Lambda^{(0)}_2(2,3) +  V   G  \Lambda_2(2,3)   G
\end{align}
where
\begin{align}\label{eq:sc_ver20}
  \Lambda^{(0)}_2(2,3) =
 -  \sum_{{\cal P}(2;3)}  V  G \Lambda_1(2)   G  \Lambda_1(3)  G~.
\end{align}
\subsubsection{Interaction induced three-photon vertex}\label{app:ver3}
The three-photon vertex function is defined as follows  
\begin{align}
\Lambda_3(2,3,4)
=\frac{1}{2}\frac{\delta^3   G^{-1}}{\delta A(2)\delta A(3)\delta A(4)} 
=\frac{1}{2}\frac{\delta^3   G^{-1}_0}{\delta A(2)\delta A(3)\delta A(4)} -\frac{1}{2}\frac{\delta^3   \Sigma}{\delta A(2)\delta A(3)\delta A(4)}~. 
\end{align}
The third derivative of the bare Green's function vanishes in Dirac systems which implies the absence of bare three-photon vertex. However, an interaction induced three-photon vertex function is obtained owing to the field-dependent self-energy.  Since the self-energy depends on the external field only through the dependence on the Green's function, we have 
\begin{align}
\Lambda_3(2,3,4)
=-\frac{1}{2}\frac{\delta^3   \Sigma}{\delta A(2)\delta A(3)\delta A(4)} 
=-\frac{1}{2} V \frac{\delta^3   G}{\delta A(4)\delta A(3)\delta A(2)}~. 
\end{align}
Using  Eq.~(\ref{eq:d3G}), we obtain 
\begin{align}\label{eq:sc_ver3}
\Lambda_3(2,3,4)= \Lambda^{(0)}_3(2,3,4) + V G \Lambda_3(2,3,4) G
\end{align}
where
\begin{align}\label{eq:sc_ver30}
\Lambda^{(0)}_3(2,3,4)
=
 \frac{1}{2}\sum_{{\cal P}(2;3;4)} 
   V\Bigg[
G  \Lambda_1(2)   G  \Lambda_1(3)  G  \Lambda_1(4)  G
- 
\frac{1}{2}   G   \Lambda_2(2,3)   G  \Lambda_1(4)  G
- 
\frac{1}{2}   G    \Lambda_1(4)  G   \Lambda_2(2,3)  G\Bigg]~.
\end{align}
To summarise, the formal derivation given in the current Section guides us in constructing a conserving diagrammatic theory for the third-order response function in Dirac systems. Lengthy mathematical relations for different contributions to the nonlinear response function, i.e. Eq.~(\ref{eq:chi3}), and the multi-photon vertex functions, i.e. Eqs.~(\ref{eq:sc_ver1}),(\ref{eq:sc_ver2}),(\ref{eq:sc_ver20}),(\ref{eq:sc_ver3}), and (\ref{eq:sc_ver30}), are graphically illustrated in Feynman diagrams depicted in Fig.~1 of the main text. Quantitative evaluation of these diagrams are explicitly discussed with great details in the next Section.  
\section{Analytical expressions for nonlinear conductivity} 
 We present here the analytical expression of the third-order optical
 response function which in the Matsubara space can be formally written as:
\begin{align}
\rchi^{(3)}_{\rm THG} (m_1,m_2,m_3) =   \frac{1}{\beta}\sum_{n} P(n,n+m_1,n+m_1+m_2,n+m_1+m_2+m_3),
\end{align}
where $n$ stands for a short notation $n=i\omega_n$ ($\omega_n$ being a fermionic frequency),
and $m=i\omega_m$ ($\omega_m$ being a bosonic frequency).
For the third-harmonic generation we have $m_1=m_2=m_3=m$ and
\begin{align}
\rchi^{(3)}_{\rm THG} (m) =   \frac{1}{\beta}\sum_{n} P(n,n+m,n+2m,n+3m).
\end{align}
After a straightforward algebra we perform the Matsubara summation and analytic continuation as $i\omega_m\to \hbar\omega+i0^+$, we obtain (see Section \ref{sec:analytical-continuation})
\begin{align}
\rchi^{(3)}_{\rm THG}(\omega) &= 
 \int^{+\infty}_{-\infty} \frac{d\epsilon}{2\pi i} 
\bigg\{ 
n_{\rm F}(\epsilon) P^{\rm RRRR}
-n_{\rm F}(\epsilon+3\hbar\omega) P^{\rm AAAA} 
+  (n_{\rm F}(\epsilon+\hbar\omega) 
-n_{\rm F}(\epsilon)  )  P^{\rm ARRR}
\nonumber\\&
+  (n_{\rm F}(\epsilon+2\hbar\omega) 
- n_{\rm F}(\epsilon+\hbar\omega) )P^{\rm AARR}
+ (n_{\rm F}(\epsilon+3\hbar\omega) 
-n_{\rm F}(\epsilon+2\hbar\omega) ) P^{\rm AAAR}
\bigg\}~.
\end{align}
Note that $P^{\rm RRRR}(\epsilon_0,\epsilon_1,\epsilon_2,\epsilon_3)$ with $\epsilon_j = \epsilon+j \omega+i\eta_j$ means that all frequency arguments are in the retarded channel (i.e. $(\eta_j \to0^+$) while $P^{\rm ARRR}$ implies that the first argument is in the advanced channel,  $\eta_0\to-0^+$, but the other are retarded, $\eta_{j\neq 0}\to0^+$. Therefore, we have $P^{\rm AAAA} = (P^{\rm RRRR})^\ast$.
By knowing the response function, the third-harmonic optical conductivity reads 
\begin{align}
\sigma^{(3)}_{\rm THG}(\omega)
=  i \frac{\rchi^{(3)}_{\rm THG}(\omega) }{\omega^3}~.
\end{align}
\subsection{Different diagram contributions}
Using Baym-Kadanoff analysis summarised in the previous section, we construct diagrams for the third-order response function in Dirac materials as illustrated in Fig.~1 of the main text. 
As depicted in Fig.~1a of the main text, the $P$-function contains four main 
different contributions, $P= P_1+P_2+P_3+P_4$,
associated respectively with square ($P_1$), triangles ($P_2$, $P_3$) and bubble ($P_4$) diagrams.
More explicitly we can write:
 \begin{align}
P_{1}(z_0,z_1,z_2,z_3) &
= \left(\frac{e^4 v^2 N_f}{2\pi \hbar^2}\right)
 Q_1(z_0,z_1)Q_1(z_1,z_2)Q_1(z_2,z_3)
 \Omega_1(z_0,z_1,z_2,z_3)
 \end{align}
The sum over spin and valley indices just leads to an overall degeneracy factor $N_f=N_s N_v$ where $N_s=2$ and $N_v=2$.  Note that $\Omega_1(z_0,z_1,z_2,z_3)$ is the bare square diagram in the absence
 of vertex renormalization,
 \begin{eqnarray}
 \Omega_1(z_0,z_1,z_2,z_3) &=\frac{\gamma_{\rm imp}}{2U}
\sum_{{\bf k}} {\rm Tr} [\hat\sigma_y \hat G({\bf k},z_0)\hat\sigma_y  \hat G({\bf k},z_1)\hat\sigma_y 
 \hat G({\bf k},z_2) \hat \sigma_y \hat G({\bf k},z_3)]
 ,
 \end{eqnarray}
and the one-photon vertex $\hat \Lambda_1(z_i,z_j) = (-ev\hat\sigma_y) \Lambda^{(0)}_1 Q_1(z_i,z_j)$ 
where  $\Lambda^{(0)}_1=1$ is the bare one-photon vertex function and $Q_1(z_i,z_j)$ is the one-photon Bethe-Salpeter renormalization factor 
which is given in the following subsection.
In similar way we can write the contributions 
 of the two triangles diagrams, namely
 \begin{align}
P_{2}(z_0,z_1,z_2,z_3) & = 
\left(\frac{e^4 v^2 N_f}{2\pi \hbar^2 }\right)  
Q_1(z_2,z_3)Q_2(z_0,z_2)
\Lambda_2^{(0)}(z_0,z_1,z_2)\Omega_2(z_0,z_2,z_3)
,
\end{align}
where
 \begin{align}
\Omega_2(z_0,z_2,z_3) &= -\frac{\gamma_{\rm imp}}{2U}
\sum_{{\bf k}}{\rm Tr} [\hat\sigma_y \hat G({\bf k},z_0) \hat G({\bf k},z_2)\hat \sigma_y \hat G({\bf k},z_3)]~,
 \end{align}
 Note that $\hat \Lambda_2(z_0,z_1,z_2) = (-ev\hat\sigma_y)^2 \Lambda_2^{(0)}(z_0,z_1,z_2) Q_2(z_0,z_2)$ is the two-photon vertex function where $\Lambda_2^{(0)}(z_0,z_1,z_2)$ is the unrenormalized two-photon vertex function and  $Q_2(z_0,z_2)$ is the two-photon Bethe-Salpeter
 renormalization factor (see subsections below).
We have also the further triangle diagram:
 \begin{align}
P_{3}(z_0,z_1,z_2,z_3) & = 
\left(\frac{e^4 v^2 N_f}{2\pi \hbar^2 }\right)  
Q_1(z_0,z_1)Q_2(z_1,z_3)
\Lambda_2^{(0)}(z_1,z_2,z_3)\Omega_3(z_0,z_1,z_3)~.
\end{align}
where
 \begin{align}
 \Omega_3(z_0,z_1,z_3) &= - \frac{ \gamma_{\rm imp}}{2U}
\sum_{{\bf k}} {\rm Tr} [\hat\sigma_y \hat G({\bf k},z_0)  \hat\sigma_y \hat G({\bf k},z_1) \hat G({\bf k},z_3)]~,
 \end{align}
Finally we make the bubble term explicit: 
\begin{align}
P_{4}(z_0,z_1,z_2,z_3) &= 
\left(\frac{e^4 v^2 N_f}{2\pi \hbar^2}\right)  \Lambda^{(0)}_3(z_0,z_1,z_2, z_3) 
Q_3(z_0,z_3) X_1(z_0,z_3)
,
\end{align}
where $\hat\Lambda_3(z_0,z_1,z_2, z_3) =(-ev\hat\sigma_y)^3\Lambda^{(0)}_3(z_0,z_1,z_2, z_3) Q_3(z_0,z_3)$ is the three-photon vertex function with $\Lambda^{(0)}_3(z_0,z_1,z_2, z_3)$ being the unrenormalized three-photon vertex function and $Q_3(z_0,z_3)$ is the three-photon Bethe-Salpeter renormalizationfactor (see subsections below). The explicit expressions of  $\Lambda^{(0)}_2(z_0,z_1,z_2)$, $\Lambda^{(0)}_3(z_0,z_1,z_2, z_3)$, and $Q_{n=1,2,3}(z_i,z_j)$, $ \Omega_1(z_0,z_1,z_2,z_3) $, $\Omega_2(z_0,z_2,z_3)$, and $\Omega_3(z_0,z_1,z_3)$ are provided in the next subsections.
 \subsection{Renormalization of the one-photon vertex}
The one-photon vertex renormalization is depicted diagrammatically in Fig.~1b of the main text and it reads
\begin{align}
 \hat\Lambda_1({\bf p},{\bf p}+{\bf q};n,n+m)
 - \hat\Lambda^{(0)}_1({\bf p},{\bf p}+{\bf q};n,n+m) 
=  \gamma_{\rm imp}  
\sum_{{\bf k}} \hat G({\bf k},n)  \hat \Lambda_1({\bf k},{\bf k}+{\bf q};n,n+m) \hat G({\bf k}+{\bf q},n+m) ,
\end{align}
where $m\equiv iq_m$ and $n \equiv ik_n=ip_n$ stand for the bosonic and fermionic Matsubara frequencies, respectively. Note that in the integrand we have shifted the dummy momentum $\bf k$ as ${\bf k} +{\bf p}\to {\bf k}$ and therefore we can  see that vertex correction does not depends on the fermion momentum $\bf p$. 
For the optical (or dipole) approximation we have $\bf q=0$.  Therefore, the Bethe-Salpeter relation for the one-photon vertex function reads   
\begin{align}
\hat\Lambda_{1}(n,n+m)= \hat \Lambda^{(0)}_1
+ \gamma_{\rm imp} \sum_{{\bf k}}
 \hat G(k,n)\hat\Lambda_{1}(n,n+m) \hat G (k,n+m)~.
\end{align}
Note that we have $ \hat \Lambda^{(0)}_1 ={\delta \hat G^{-1}_0}/{\delta A}|_{\bf A\to 0} = -ev \hat\sigma_y$ where $\hat G_0$ stands for the non-interacting Green's function. 
We assume the following ansatz for the vertex function 
\begin{align}
\hat \Lambda_1(n,n+m) = a\hat I+b\sigma_x+(c+v)\sigma_y+d\sigma_z~.
\end{align}
Using the fact that the integral of odd-function of ${\bf k}$ is zero, we obtain $a=b=d=0$ and eventually the following result for the vertex function $\hat \Lambda_1 =  (-ev \hat\sigma_y) \Lambda_1$ where
\begin{align}
 \Lambda_1(n,n+m)  = \frac{ \Lambda^{(0)}_1}{1- U X_1(n,n+m) }~. 
 \label{eql1}
\end{align}
Note that $\Lambda^{(0)}_1=1$ and 
\begin{align}
 X_1(n,n+m) = \frac{\gamma_{\rm imp}}{2 U} \sum_{\bf k} {\rm Tr}[\hat\sigma_y \hat G({\bf k},n)\hat\sigma_y \hat G({\bf k},n+m)]~.
\end{align}
Using dimensional regularization, we find the following formula for the $X_1$ function: 
\begin{align}
X_1(n,n+m)  =  \frac{S(n)S(n+m)}{S(n)^2-S(n+m)^2}\ln\left[\frac{S(n+m)^2}{S(n)^2}\right]~.
\end{align}
The above equation can be straightforwardly generalized in the generic complex space by replacing
$n \to i\omega_n \to z_1$, $n+m \to i\omega_n+i\omega_m \to z_2$. Eq. (\ref{eql1}) defines the one-photon vertex renormalization factor:
\begin{align}
 Q_1(z_1,z_2)  = \frac{1}{1- U X_1(z_1,z_2) }~. 
\end{align}
Obviously $X_1(z_1,z_2) = X_1(z_2,z_1)$. 
\subsection{Renormalization of the two-photon vertex}
Similar to the one-photon vertex case, it can be shown that the two-photon vertex function is independent of the fermionic momentum, $\bf p$. Moreover, for the optical limit we can neglect the photon momentum $\bf q$. 
The self-consistent Bethe-Salpeter relation for the two-photon vertex function is depicted in Fig.~1c of the main text and it reads 
\begin{align}\label{eq:BS_2photon}
\hat \Lambda_{2}(n,n+m,n+2m)  = 
\hat {\Lambda}^{(0)}_{2} (n,n+m,n+2m)
+ \gamma_{\rm imp}  
\sum_{\bf k} \hat G({\bf k},n)  \hat \Lambda_{2} (n,n+m,n+2m) \hat G({\bf k},n+2m)~. 
\end{align}
In the non-interacting Dirac system the ``bare''    
two-photon vertex function is zero, $\hat \Lambda^{(0)}_{2} \propto \delta^2 \hat G^{-1}_0/\delta A^2|_{\bf A\to 0}=0$, due to the linear momentum dependence of the Hamiltonian. However, due to interaction the unrenormalized two-photon vertex $\hat \Lambda^{(0)}_{2}$ is finite given by the following relation (see Fig.~1e of the main text) 
\begin{align}
&\hat {\Lambda}^{(0)}_{2} (n,n+m,n+2m) = 
-\gamma_{\rm imp}  
\sum_{{\bf k}} \hat G({\bf k},n) 
\hat \Lambda_y(n,n+m) 
\hat G({\bf k},n+m)
\hat \Lambda_y(n+m,n+2m)  
\hat G({\bf k},n+2m)~. 
\end{align} 
From now on we adopt the short-hand notation $z_j = n+j m$ with $j=0,1,2,3$. 
We find $\hat \Lambda^{(0)}_{2} =  (-ev\hat\sigma_y)^2  \Lambda^{(0)}_2$ with $\hat\sigma^2_y=\hat I$
and 
\begin{align}
\Lambda^{(0)}_2(z_0,z_1,z_2) =  
Q_1(z_0,z_1)Q_1(z_1,z_2)U   Z(z_0,z_1,z_2)
,
\end{align} 
in which $Q_1(z_i,z_j)$ is the one-photon renormalization factor defined in the previous subsection,
and where
\begin{align}
 Z(z_0,z_1,z_2) = -\frac{\gamma_{\rm imp }}{2 U} \sum_{{\bf k}} {\rm Tr}[\hat G({\bf k},z_0) \hat \sigma_y 
\hat G({\bf k},z_1)
\hat \sigma_y  \hat G({\bf k},z_2)]~.
\end{align}
By performing the momentum integration using the dimensional regularization, we obtain 
\begin{align}
  Z(z_0,z_1,z_2) & = \frac{S(z_1)}{S(z_0)-S(z_2)} \Bigg\{ 
\frac{S(z_0)  }{S(z_0)^2-S(z_1)^2}
 \ln\left[\frac{S(z_1)^2}{S(z_0)^2}\right]
 + \frac{ S(z_2)}{S(z_1)^2-S(z_2)^2}
  \ln\left[\frac{S(z_1)^2}{S(z_2)^2}\right]
 \Bigg\}~.
\end{align} 
In a compact form we can write 
\begin{align}
Z(z_0,z_1,z_2)  = \frac{X_1(z_0,z_1) - X_1(z_1,z_2)}{S(z_0)-S(z_2)}~.
\end{align}
By solving the the self-consistent Bethe-Salpeter relation for the two-photon vertex given in Eq.~(\ref{eq:BS_2photon}), we obtain $\hat \Lambda_{2}= (-ev\hat\sigma_y)^2 \Lambda_2$  with   
\begin{align}
  \Lambda_{2}(z_0,z_1,z_2)  = 
  Q_2(z_0,z_2) 
  \Lambda^{(0)}_{2}(z_0,z_1,z_2)
  ,
  \end{align}
  in which $Q_2(z_0,z_2)$ is the two-photon Bethe-Salpeter renormalization factor
 \begin{align}
 Q_2(z_0,z_2) 
 =
 \frac{1}
{1-U X_2(z_0,z_2)},
\end{align}
 and
where  
\begin{align}
X_2(z,z') = \frac{\gamma_{\rm imp}}{2 U} \sum_{\bf k} {\rm Tr}[\hat G({\bf k},z) \hat G({\bf k},z')].
\end{align}
We explicitly obtain 
\begin{align}
X_2(z,z') 
= \frac{1}{S(z)-S(z')}\Bigg\{ S(z)\ln\left[-\frac{W^2}{S(z)^2}\right]
-
S(z') \ln\left[-\frac{W^2}{S(z')^2}\right] 
\Bigg\}~.
\end{align}
Note that $X_2(z,z') = X_2(z',z)$. Using the self-energy relation Eq.~(\ref{eq:self}) we have 
\begin{align}
S(z)\ln\left[-\frac{W^2}{S(z)^2}\right] = -\frac{\Sigma(z)}{U}~.
\end{align}
Therefore, the two-photon renormalizationfactor reads (Eq.~(6) of the main text)
\begin{align}
Q_2(z,z') = \frac{1}{1-U X_2(z,z')}  = \frac{S(z)-S(z')}{z-z'}~.
\end{align}
It is useful to evaluate  of $Q^{\rm RR}_2(\epsilon,\epsilon)$ which is given in Eq. (7) of the main text. 
Using the above relation in  the retarded-retarded (RR) channel and employing the self-energy relation Eq.~(\ref{eq:self}), we obtain  
\begin{align}
Q^{\rm RR}_2(\epsilon,\epsilon) &= S'(\epsilon)  = \frac{d  S(\epsilon) }{d\epsilon}
= 1  - \frac{d\Sigma(\epsilon)}{d\epsilon}
\nonumber\\&
= 1 + U S'(\epsilon) \ln \left[ - \frac{W^2}{S(\epsilon)^2}\right ]
 - U S(\epsilon) \frac{d \ln \left[S(\epsilon)^2\right ]}{d\epsilon}
 \nonumber\\&
= 1 + U S'(\epsilon) \ln \left[ - \frac{W^2}{S(\epsilon)^2}\right ]
 - 2U S'(\epsilon)~. 
\end{align}
Therefore, we have 
\begin{align}
\left [1 + 2U - U  \ln \left[ - \frac{W^2}{S(\epsilon)^2}\right ] \right] S'(\epsilon) = 1~.
\end{align}
Consequently, we arrive at 
\begin{align}
S'(\epsilon) = \frac{1}{1 + 2U - U  \ln \left[ - \frac{W^2}{S(\epsilon)^2}\right ]}~.
\end{align}
Using  Eq.~(\ref{eq:self}), we obtain  
\begin{align}
1-U  \ln \left[-\frac{W^2}{S(\epsilon)^2}\right ] = 1+\frac{\Sigma(\epsilon)}{S(\epsilon)} =\frac{\mu_0+\epsilon}{S(\epsilon)}~.
\end{align}
Therefore, we find the following relation which is given in Eq.~(7) of the main text. 
\begin{align}
Q^{\rm RR}_2(\epsilon,\epsilon) = S'(\epsilon) = \frac{S(\epsilon)}{ 2U S(\epsilon) +\mu_0+\epsilon}~.
\end{align}
\subsection{Renormalization of the three-photon vertex}
Similar to the case of two-photon case, the impurity scattering induces a finite three-photon vertex as defined in Fig.~1f of the main text.
Accordingly we find  $\hat{\Lambda}^{(0)}_{2}= (-ev\hat \sigma_y)^3  \Lambda^{(0)}_3$ with 
\begin{align}
\Lambda^{(0)}_3(z_0,z_1,z_2,z_3)  =   M_1(z_0,z_1,z_2,z_3) + M_2(z_0,z_1,z_2,z_3) + M_3(z_0,z_1,z_2,z_3)
\end{align}
where
\begin{align}
& M_1(z_0,z_1,z_2,z_3) 
 =
 U \Omega_1(z_0,z_1,z_2,z_3)Q_1(z_0,z_1)Q_1(z_1,z_2)Q_1(z_2,z_3)
 ,
\\
& M_2(z_0,z_1,z_2,z_3)  =
U  \Omega_2(z_0,z_2,z_3) \Lambda^{(0)}_2 (z_0,z_1,z_2) Q_1(z_2,z_3)Q_2(z_0,z_2)
,
\\
& M_3(z_0,z_1,z_2,z_3) =
U \Omega_3(z_0,z_1,z_3) \Lambda^{(0)}_2(z_1,z_2,z_3)Q_1(z_0,z_1)Q_2(z_1,z_3)
.
 \end{align}
Here $Q_1(z_i,z_j)$, $Q_2(z_i,z_j)$ are the Bethe-Salpeter one- and two-photon renormalization functions, 
respectively.
The explicit expression for  $\Omega_1$ function is given by   
\begin{align}
 \Omega_1(z_0,z_1,z_2,z_3)
 =
 u_1(z_0,z_1,z_2,z_3)  \ln\left[\frac{S(z_0)^2}{S(z_1)^2}\right]
  +
 u_2(z_0,z_1,z_2,z_3) \ln\left[\frac{S(z_0)^2}{S(z_2)^2}\right]
 +
 u_3(z_0,z_1,z_2,z_3)  \ln\left[\frac{S(z_0)^2}{S(z_3)^2}\right]
\end{align}
where
\begin{align}
&u_1(z_0,z_1,z_2,z_3)= \frac{S(z_0)S(z_1)S(z_2)S(z_3)+ [S(z_0)S(z_2)+S(z_1)S(z_3)]S(z_1)^2}
{[S(z_1)^2-S(z_0)^2][S(z_1)^2-S(z_2)^2][S(z_1)^2-S(z_3)^2]}~,
\\
&u_2(z_0,z_1,z_2,z_3)= \frac{S(z_0)S(z_1)S(z_2)S(z_3)+ [S(z_0)S(z_2)+S(z_1)S(z_3)]S(z_2)^2}
{[S(z_2)^2-S(z_0)^2][S(z_2)^2-S(z_1)^2][S(z_3)^2-S(z_3)^2]}~,
\\
&u_3(z_0,z_1,z_2,z_3)= \frac{S(z_0)S(z_1)S(z_2)S(z_3)+ [S(z_0)S(z_2)+S(z_1)S(z_3)]S(z_3)^2}
{[S(z_3)^2-S(z_0)^2][S(z_3)^2-S(z_1)^2][S(z_3)^2-S(z_2)^2]}~.
\end{align}
Similarly, one can obtain  
\begin{align}
\Omega_2(z_0,z_2,z_3) = Z(z_2,z_3,z_0)~,
\\
\Omega_3(z_0,z_1,z_3) = Z(z_3,z_0,z_1)~.
\end{align}
Finally, the Bethe-Salpeter renormalization of the three-photon vertex function gives:
\begin{align}
 \Lambda_{3}(z_0,z_1,z_2,z_3) = Q_3(z_0,z_3) \Lambda^{(0)}_{3}(z_0,z_1,z_2,z_3)~.
\end{align}
where 
\begin{align}
Q_3 (z_1,z_2) = \frac{1}{1-U X_3(z_1,z_2)}
\end{align}
Note that $X_3(z_1,z_2)=X_1(z_1,z_2)$. 
\section{Linear conductivity} 
\begin{figure}[h]
\centering
\begin{overpic}[width=80mm]{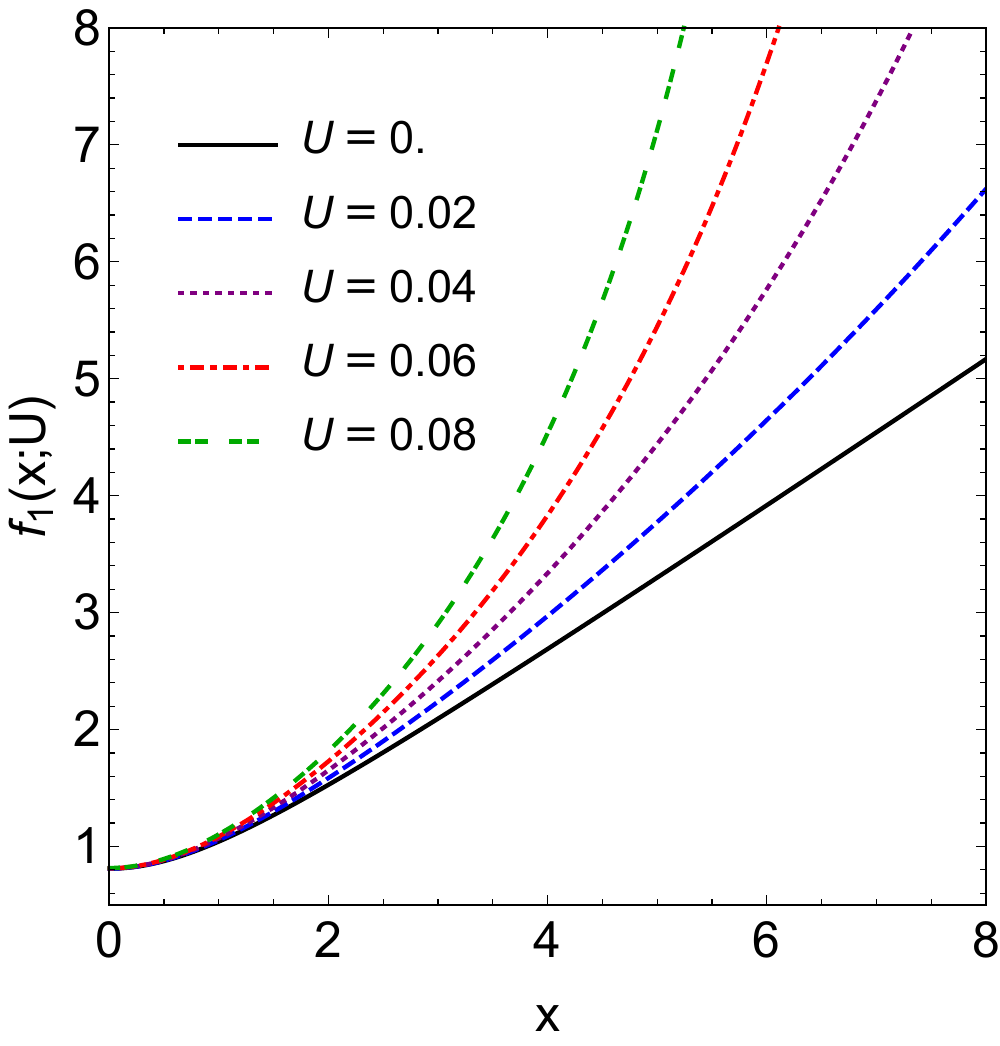}\end{overpic}
\caption{Universal $f_1(x;U)$ function versus $x$ for several values of $U$.}
\label{fig:f1}
\end{figure}

Linear response function is obtained after performing a Matsubara summation as follows
\begin{align}
\rchi^{(1)}(m) = \frac{1}{\beta} \sum_n P(n,n+m)~. 
\end{align}
Note that $\beta=1/k_BT$ where $k_B$ is the Boltzmann constant and $T$ stands for the electronic temperature. 
The $P$-function is analytical in the complex plain except two branch cuts at $\epsilon$ and $\epsilon-m$ where $\epsilon$ span over whole real axes. After performing the summation and an analytical continuation as $m\to \hbar\omega+i0^+$, we find \cite{mahan}
\begin{align}
\rchi^{(1)}(\omega) = \int^{\infty}_{-\infty} \frac{d\epsilon}{2\pi i} 
\left\{  [n_{\rm F}(\epsilon)-n_{\rm F}(\epsilon+\hbar\omega)]P^{\rm AR}(\epsilon,\epsilon+\hbar\omega)
 -n_{\rm F}(\epsilon)P^{\rm RR}(\epsilon,\epsilon+\hbar\omega)+ n_{\rm F}(\epsilon+\hbar\omega)P^{\rm AA}(\epsilon,\epsilon+\hbar\omega)
 \right\} 
\end{align}
where $n_{\rm F}(x) =1/(e^{\beta x}+1)$ is the Fermi-Dirac distribution function. 
Note that ``R'' and ``A'' superscripts stand for the retarded and advanced, respectively.
Accordingly, we have $P^{\rm RR}(\epsilon,\epsilon+\hbar\omega)=[P^{\rm AA}(\epsilon,\epsilon+\hbar\omega)]^\ast= P(\epsilon+i0^+,\epsilon+\hbar\omega+i0^+)$ and $P^{\rm AR}(\epsilon,\epsilon+\hbar\omega)=P(\epsilon-i0^+,\epsilon+\hbar\omega+i0^+)$. 
The linear optical conductivity reads
\begin{align}
\sigma^{(1)}(\omega)
=  i \frac{\rchi^{(3)}(\omega) }{\omega}~.
\end{align}
For the 2D Dirac model, the $P$-function reads 
\begin{align}
P(n,n+m) = \frac{N_f e^2 }{2\pi \hbar^2}  \frac{X_1(n,n+m)}{1-UX_1(n,n+m)}~.
\end{align}
Note that the sum over spin and valley index just leads to an overall degeneracy factor $N_f=N_s N_v$ where $N_s=2$ and $N_v=2$. 
The linear $dc$-conductivity follows 
\begin{align}
\sigma^{(1)}_{\rm dc} =  {\frac{4  \sigma_0}{\pi^2}} \left\{\frac{X^{\rm AR}_1(0,0)}{1-UX^{\rm AR}_1(0,0)}
 -\frac{X^{\rm RR}_1(0,0)}{1-UX^{\rm RR}_1(0,0)}\right \} 
\end{align}
where $\sigma_0 = e^2/4 \hbar$. 
The retarded self-energy can be decomposed into its real and imaginary parts $\Sigma(\epsilon) = \Delta(\epsilon) - i\Gamma(\epsilon)$ with $\Delta(\epsilon)$ and $\Gamma(\epsilon)>0$ being odd and even real functions, respectively. Using this notation we find $X^{\rm RR}_1(0,0) =-1$,  
$X^{\rm AR}_1(0,0)=w(x)$ with  $x = \mu/\Gamma(\mu)$ in which the renormalized chemical potential is given by $\mu= \mu_0-\Delta(\mu)$, and  $w(x)$ reads 
\begin{align}
w(x)= \frac{1+x^2}{2x}
{\rm arctan}\left[ \frac{1-x^2}{1+x^2},
\frac{2x}{1+x^2} \right]~.
\end{align}
Therefore, we find
\begin{align}
\sigma^{(1)}_{\rm dc} = \sigma_0 f_1\left(\frac{\mu}{\Gamma(\mu)}; U\right) 
\end{align} 
where
\begin{align}
f_1(x; U) = {\frac{4 }{\pi^2}}   \left\{\frac{1}{1+U}+\frac{w(x)}{1-U w(x)}\right \}~. 
\end{align}
The functional dependence of $f_1(x;U)$ on $x$ and $U$ is illustrated in Fig.~\ref{fig:f1}. 
In the constant-$\Gamma$ model, $\Sigma = - i\Gamma$, we have 
\begin{align}
f_1(x) =\frac{4  [1+w(x)]  }{\pi^2}~.
\end{align}
The asymptotic form of $f_1(x)$ for small and large $x$ follows
\begin{align}
&f_1(x) \approx \frac{8}{\pi^2} \left(1+\frac{x^2}{3}\right)~~~,~~~x\ll 1~,
\\
&f_1(x) \approx \frac{2}{\pi} \left(x + \frac{1}{ x}\right)~~~~~~,~~~x\gg 1~.
\end{align}

\section{Nonlinear dc conductivity} 
\subsection{Analytical derivation $f_3(x)$ function in the constant-$\Gamma$ model }
Nonlinear dc conductivity is given by 
\begin{align}
\sigma^{(3)}_{\rm dc} = - \lim_{\omega\to 0} \frac{{\rm Im}[\rchi^{(3)}_{\rm THG}(\omega)] }{\omega^3}
\end{align}
where the imaginary part of nonlinear (third-harmonic) response function follows
\begin{align}
 {\rm Im}[\rchi^{(3)}_{\rm THG} (\omega)] & = \int^{+\infty}_{-\infty} \frac{d\epsilon}{2\pi } 
\Big\{ \left (n_{\rm F}(\epsilon+3\hbar\omega)-n_{\rm F}(\epsilon)  \right ) {\rm Re}[P^{\rm RRRR}(\epsilon,\epsilon+\hbar\omega, \epsilon+2\hbar\omega,\epsilon+3\hbar\omega)]
\nonumber\\&-  (n_{\rm F}(\epsilon+\hbar\omega) -n_{\rm F}(\epsilon)  )  {\rm Re}[P^{\rm ARRR}(\epsilon,\epsilon+\hbar\omega, \epsilon+2\hbar\omega,\epsilon+3\hbar\omega)]
\nonumber\\&
-  (n_{\rm F}(\epsilon+2\hbar\omega) - n_{\rm F}(\epsilon+\hbar\omega) ){\rm Re}[P^{\rm AARR}(\epsilon,\epsilon+\hbar\omega, \epsilon+2\hbar\omega,\epsilon+3\hbar\omega)]
\nonumber\\&- (n_{\rm F}(\epsilon+3\hbar\omega) -n_{\rm F}(\epsilon+2\hbar\omega) ) {\rm Re}[P^{\rm AAAR}(\epsilon,\epsilon+\hbar\omega, \epsilon+2\hbar\omega,\epsilon+3\hbar\omega)]
\Big\}~.
\end{align}
For small frequency we can expand the Fermi function as follows   
\begin{align}
n_{\rm F}(\epsilon+\hbar\omega) -n_{\rm F}(\epsilon)  &\approx  
\hbar\omega \partial_\epsilon n_{\rm F}(\epsilon) + \frac{(\hbar\omega)^2}{2}
\partial^2_\epsilon n_{\rm F}(\epsilon) + \frac{(\hbar\omega)^3}{6}
\partial^3_\epsilon n_{\rm F}(\epsilon) +\dots 
\nonumber\\& = 
-(\hbar\omega)\delta(\epsilon) -  \frac{(\hbar\omega)^2}{2} \partial_\epsilon \delta(\epsilon) -  \frac{(\hbar\omega)^3}{6} \partial^2_\epsilon \delta(\epsilon) + \dots ~.
\end{align}
Notice that at zero temperature we have $n_{\rm F}(\epsilon) = \Theta(-\epsilon)$ that is the Heaviside step function and its first derivative $\partial_\epsilon n_{\rm F}(\epsilon) = -\delta(\epsilon) $ which stands for the Dirac delta function. After integration over $\epsilon$, we find 
\begin{align}
 {\rm Im}[\rchi^{(3)}_{\rm THG} (\omega)] & \approx - (\hbar\omega) h_1(\omega) + \frac{(\hbar\omega)^2}{2} h_2(\omega) - \frac{(\hbar\omega)^3}{6} h_3(\omega)
\end{align}
where we define 
\begin{align}
h_1(\omega)&=3 {\rm Re}[P^{\rm RRRR}(0,\hbar\omega, 2\hbar\omega,3\hbar\omega)]
-  {\rm Re}[P^{\rm ARRR}(0,\hbar\omega, 2\hbar\omega,3\hbar\omega)]
\nonumber\\&-{\rm Re}[P^{\rm AARR}(0,\hbar\omega, 2\hbar\omega,3\hbar\omega)] 
 - {\rm Re}[P^{\rm AAAR}(0,\hbar\omega, 2\hbar\omega,3\hbar\omega)]~,
\end{align}
\begin{align}
h_2(\omega)&=\lim_{\epsilon\to 0} \frac{\partial}{\partial\epsilon}\Big\{9 {\rm Re}[P^{\rm RRRR}(\epsilon,\epsilon+\hbar\omega, \epsilon+2\hbar\omega,\epsilon+3\hbar\omega)]
-  {\rm Re}[P^{\rm ARRR}(\epsilon,\epsilon+\hbar\omega, \epsilon+2\hbar\omega,\epsilon+3\hbar\omega)]
\nonumber\\&-3{\rm Re}[P^{\rm AARR}(\epsilon,\epsilon+\hbar\omega, \epsilon+2\hbar\omega,\epsilon+3\hbar\omega)] 
 -5 {\rm Re}[P^{\rm AAAR}(\epsilon,\epsilon+\hbar\omega, \epsilon+2\hbar\omega,\epsilon+3\hbar\omega)]
 \Big\}~,
\end{align}
\begin{align}
h_3(\omega)&=\lim_{\epsilon\to 0} \frac{\partial^2}{\partial\epsilon^2}\Big\{27 {\rm Re}[P^{\rm RRRR}(\epsilon,\epsilon+\hbar\omega, \epsilon+2\hbar\omega,\epsilon+3\hbar\omega)]
-  {\rm Re}[P^{\rm ARRR}(\epsilon,\epsilon+\hbar\omega, \epsilon+2\hbar\omega,\epsilon+3\hbar\omega)]
\nonumber\\&-7{\rm Re}[P^{\rm AARR}(\epsilon,\epsilon+\hbar\omega, \epsilon+2\hbar\omega,\epsilon+3\hbar\omega)] 
 -19 {\rm Re}[P^{\rm AAAR}(\epsilon,\epsilon+\hbar\omega, \epsilon+2\hbar\omega,\epsilon+3\hbar\omega)]
 \Big\}~.
\end{align}
In the absence of vertex correction, we have
\begin{align}
 P(z_0,z_1,z_2,z_3) = \left( \frac{e^4 v^2 N_f }{2\pi\hbar^2 }\right)  \Omega_1(z_0,z_1,z_2,z_3)~.  
\end{align}
We assume $\Sigma(\epsilon) = \mp i\Gamma$ where $\Gamma>0$ is a phenomenological constant and $-/+$ stands for the retarded (R) and advanced (A) channel, respectively.
For the case of $\hbar\omega\ll\mu,\Gamma$, it is legitimate to expand the integrand for small $\omega$. 
In the constant-$\Gamma$ model, the contribution from $h_2(\omega)$ exactly cancels that of $h_3(\omega)$. Eventually, the nonlinear dc conductivity in the constant-$\Gamma$ model reads
\begin{align}
\sigma^{(3)}_{\rm dc} = \left( \frac{e^4 v^2 N_f  \hbar^3 }{4\pi^2\hbar^2 t^4_0}\right) \frac{t^4_0}{\Gamma^4} f_3\left (\frac{\mu}{\Gamma}\right)~.
\end{align}
Considering $\hbar v =\sqrt{3}t_0 a/2$ and $N_f=4$, we have
\begin{align}
\left(\frac{e^4 v^2 N_f \hbar^3}{(2\pi)^2 \hbar^2 t^4_0 }\right) = \frac{\sigma_0}{E^2_0}
\end{align}
with $\sigma_0 =e^2/4\hbar$ and 
\begin{align}
E_0 = \frac{\pi}{\sqrt{3}}\frac{t_0}{e a} 
\end{align}
where with $t_0\approx 3$eV and $a\approx0.246$nm we find $E_0\approx 22 {\rm V/nm}$. The universal $f_3(x)$ function reads 
\begin{align}\label{eq:f3}
f_3(x) = \frac{3}{16} \left[\frac{1}{x^4}-\frac{1}{x^2} \right]
+\frac{2}{3} \left[\frac{1}{1+x^2} \right]^3
+\frac{1}{16} \left [ \frac{5}{x^2} - \frac{8}{1+x^2} -\frac{3}{x^4} \right]
w(x)~.
\end{align}
There is a sign change in $f_3(x)$ at $x_0 \approx 0.655$. 
It is good to check the following asymptotic cases : 
\begin{align}
&f_3(x) \approx \frac{2}{5} \left(1 - \frac{33}{7}  x^2 \right)~~~~~,~~~~~x\ll 1~,
\\
&f_3(x) \approx -\frac{3\pi}{32} \left(\frac{1}{x} - \frac{2}{3 x^3} \right)~~~,~~~x\gg1~.
\end{align}
\subsection{Numerical evaluation of $f_3(x,U)$ function in the full quantum theory} 
The third-harmonic generation (THG) conductivity is given by 
\begin{align}
\sigma^{(3)}_{\rm THG}(\omega)
&= i \frac{\rchi^{(3)}_{\rm THG}(\omega)}{\omega^3}
= \left(\frac{e^4 v^2 N_f}{2\pi \hbar^2}\right) \frac{1}{\omega^3} \int^{\infty}_{-\infty} \frac{d\epsilon}{2\pi} 
~{\rm Re}[K(\epsilon,\omega)]
\nonumber\\
&= \left( \frac{e^4 v^2 N_f \hbar^3}{(2\pi)^2 \hbar^2 t^4_0 } \right)
\left[\frac{t^4_0}{(\hbar\omega)^3} \int^{\infty}_{-\infty} d\epsilon ~{\rm Re}[K(\epsilon,\omega)]\right]
\end{align}
where 
\begin{align}
K(\epsilon,\omega) &= \left(\frac{e^4 v^2 N_f}{2\pi \hbar^2} \right)^{-1}\Big \{ n_{\rm F}(\epsilon) P^{\rm RRRR}
-n_{\rm F}(\epsilon+3\hbar\omega) P^{\rm AAAA} 
+  (n_{\rm F}(\epsilon+\hbar\omega) 
-n_{\rm F}(\epsilon)  )  P^{\rm ARRR}
\nonumber\\&
+  (n_{\rm F}(\epsilon+2\hbar\omega) 
- n_{\rm F}(\epsilon+\hbar\omega) )P^{\rm AARR}
+ (n_{\rm F}(\epsilon+3\hbar\omega) 
-n_{\rm F}(\epsilon+2\hbar\omega) ) P^{\rm AAAR}
\Big\}~.
\end{align}
Therefore, we obtain 
\begin{align}
\sigma^{(3)}_{\rm THG}(\omega)
=   \frac{\sigma_0}{E^2_0}  \frac{t^4_0}{(\hbar\omega)^3} \int^{\infty}_{-\infty} d\epsilon~
{\rm Re}[K(\epsilon,\omega)]~.
\end{align}
We evaluate the third-order dc conductivity as dc limit of the third-harmonic conductivity,
\begin{align}
\sigma^{(3)}_{\rm dc} = \lim_{\omega\to 0} \sigma^{(3)}_{\rm THG}(\omega)
=   \frac{\sigma_0}{E^2_0}  \left[\frac{t_0}{\Gamma(\mu)}\right]^4  f_3\left(\frac{\mu}{\Gamma(\mu)};U\right) 
\end{align}
where the universal $f_3$ function can be evaluated numerically  by using the following relation 
\begin{align}
f_3\left(\frac{\mu}{\Gamma(\mu)};U\right)  =\lim_{\omega\to 0} \frac{\Gamma(\mu)^4}{(\hbar\omega)^3} 
 \int^{\infty}_{-\infty} d\epsilon~{\rm Re}[K(\epsilon,\omega)]~.
\end{align}
\section{Analytical continuation for the third order response function}\label{sec:analytical-continuation}
\begin{figure}[t]
\centering
\begin{overpic}[width=120mm]{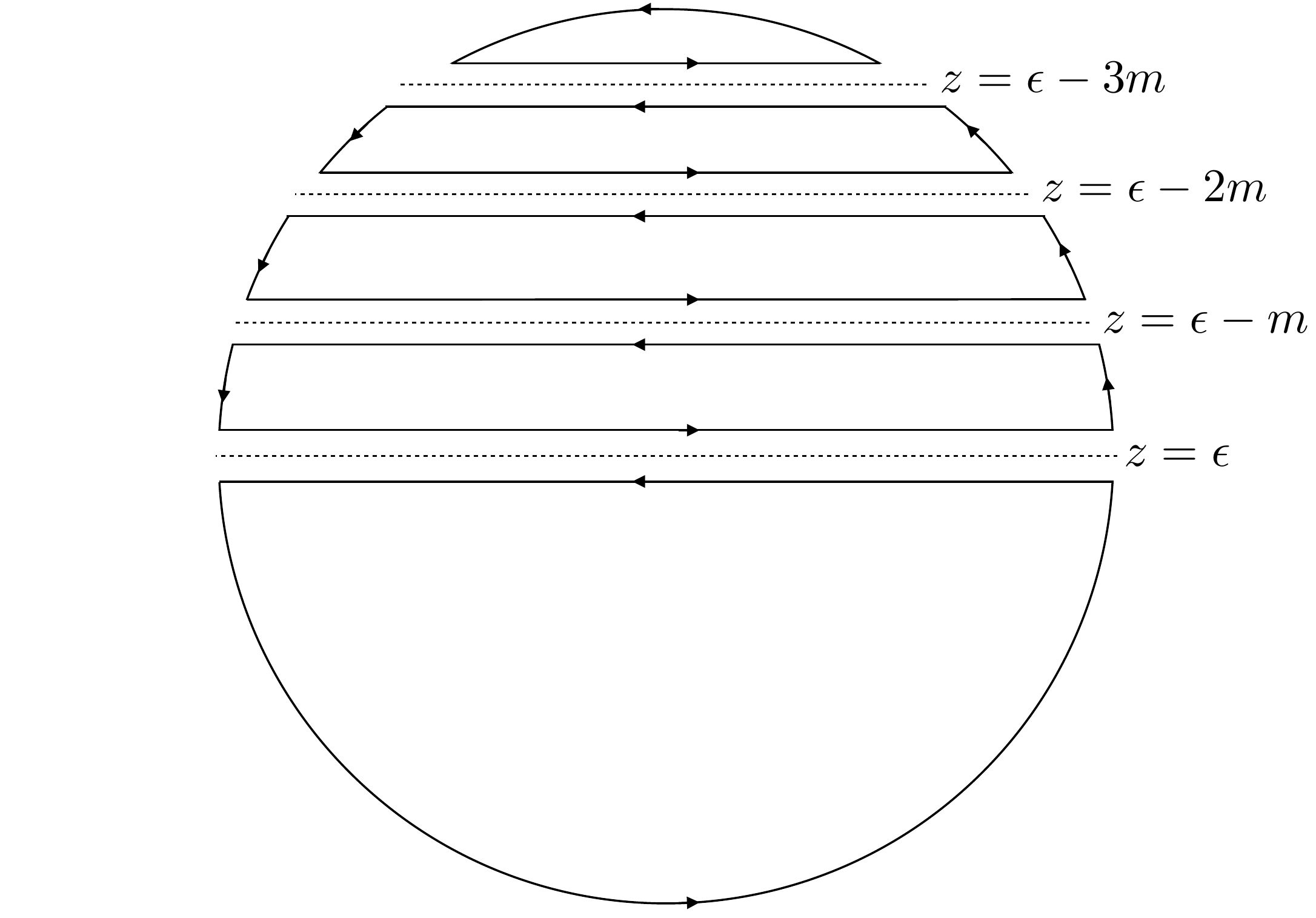}\end{overpic}
\caption{The Matsubara summation is performed by utilizing an integration on a contour enclosing whole complex plane except four branch cuts which are shown by dashed lines on the contour with radius $R\to \infty$.}
\label{fig:contour}
\end{figure}

The summation of the fermionic Matsubara frequency $n$ is performed by the contour integration technique.
\begin{align}
 B(m) 
= \frac{1}{\beta}\sum_{n} P(n , n +m, n +2m, n +3m)~.
\end{align}
The $P$-function contains four brach cut in  complex plane. The Matsubara summation is performed on a contour with four cuts at $\epsilon$, $\epsilon-m$, $\epsilon-2m$, and $\epsilon-3m$, see Fig.~\ref{fig:contour}. Note that $\epsilon$ runs over the entire real axes.
Therefore, we write 
\begin{align}
 B(m) 
&= \oint \frac{dz}{2\pi i} n_{\rm F}(z) P(z,z+m, z+2m,z+3m)
\nonumber\\
&=
\int^{+\infty}_{-\infty} \frac{d\epsilon}{2\pi i} \left \{ n_{\rm F}(\epsilon+i\eta) 
P(\epsilon+i\eta,\epsilon+i\eta+m, \epsilon+i\eta+2m,\epsilon+i\eta+3m)
- (\eta \to -\eta) \right \} 
\nonumber\\&
+ \int^{+\infty}_{-\infty} \frac{d\epsilon}{2\pi i} 
\left\{ n_{\rm F}(\epsilon-m+i\eta) 
P(\epsilon-m+i\eta,\epsilon+i\eta , \epsilon +i\eta +m,\epsilon +i\eta +2m)
- (\eta \to -\eta) \right \} 
\nonumber\\&
+\int^{+\infty}_{-\infty} \frac{d\epsilon}{2\pi i} \left \{ n_{\rm F}(\epsilon-2m+i\eta) 
P(\epsilon-2m+i\eta,\epsilon-m+i\eta, \epsilon+i\eta,\epsilon+i\eta+m)
- (\eta \to -\eta) \right \} 
\nonumber\\& 
+\int^{+\infty}_{-\infty} \frac{d\epsilon}{2\pi i} \left \{ n_{\rm F}(\epsilon-3m+i\eta) 
P(\epsilon-3m+i\eta,\epsilon-2m+i\eta, \epsilon-m+i\eta,\epsilon+i\eta)
- (\eta \to -\eta) \right \} 
\end{align}
where $n_{\rm F}(z)=1/(1+e^{\beta z})$ is the Fermi-Dirac distribution function. 
Note that $m$ stands for external bosonic Matsobara (imaginary) frequency. Since $\eta\ll |m|$, we have $m+i\eta \to m$ which implies 
\begin{align}
 B(m) &=
\int^{+\infty}_{-\infty} \frac{d\epsilon}{2\pi i} \left \{ n_{\rm F}(\epsilon+i\eta) 
P(\epsilon+i\eta,\epsilon+m, \epsilon+2m,\epsilon+3m)
- (\eta \to -\eta) \right \} 
\nonumber\\&
+ \int^{+\infty}_{-\infty} \frac{d\epsilon}{2\pi i} 
\left\{ n_{\rm F}(\epsilon-m) 
P(\epsilon-m,\epsilon+i\eta , \epsilon +m,\epsilon +2m)
- (\eta \to -\eta) \right \} 
\nonumber\\&
+\int^{+\infty}_{-\infty} \frac{d\epsilon}{2\pi i} \left \{ n_{\rm F}(\epsilon-2m) 
P(\epsilon-2m,\epsilon-m, \epsilon+i\eta,\epsilon+m)
- (\eta \to -\eta) \right \} 
\nonumber\\& 
+\int^{+\infty}_{-\infty} \frac{d\epsilon}{2\pi i} \left \{ n_{\rm F}(\epsilon-3m) 
P(\epsilon-3m,\epsilon-2m, \epsilon-m,\epsilon+i\eta)
- (\eta \to -\eta) \right \}~.
\end{align}
Note that for bosonic frequency $m$ we have $n_{\rm F}(\epsilon-m) = n_{\rm F}(\epsilon)$ and therefore we find 
\begin{align}
 B(m) &=
\int^{+\infty}_{-\infty} \frac{d\epsilon}{2\pi i} n_{\rm F}(\epsilon) \left \{ 
P(\epsilon+i\eta,\epsilon+m, \epsilon+2m,\epsilon+3m)
- (\eta \to -\eta) \right \} 
\nonumber\\&
+ \int^{+\infty}_{-\infty} \frac{d\epsilon}{2\pi i} n_{\rm F}(\epsilon)
\left\{  
P(\epsilon-m,\epsilon+i\eta , \epsilon +m,\epsilon +2m)
- (\eta \to -\eta) \right \} 
\nonumber\\&
+\int^{+\infty}_{-\infty} \frac{d\epsilon}{2\pi i} n_{\rm F}(\epsilon) \left \{ 
P(\epsilon-2m,\epsilon-m, \epsilon+i\eta,\epsilon+m)
- (\eta \to -\eta) \right \} 
\nonumber\\& 
+\int^{+\infty}_{-\infty} \frac{d\epsilon}{2\pi i} n_{\rm F}(\epsilon) \left \{  
P(\epsilon-3m,\epsilon-2m, \epsilon-m,\epsilon+i\eta)
- (\eta \to -\eta) \right \}~. 
\end{align}
We do analytical continuation as $m \to \hbar \omega + i\eta$:
\begin{align}
 B(\omega) &=
\int^{+\infty}_{-\infty} \frac{d\epsilon}{2\pi i} n_{\rm F}(\epsilon) \left \{ 
P(\epsilon+i\eta,\epsilon+\hbar\omega+i\eta, \epsilon+2\hbar\omega+i2\eta,\epsilon
+3\hbar\omega+i3 \eta)
- (\eta \to -\eta) \right \} 
\nonumber\\&
+ \int^{+\infty}_{-\infty} \frac{d\epsilon}{2\pi i} n_{\rm F}(\epsilon)
\left\{  
P(\epsilon-\hbar\omega-i\eta,\epsilon+i\eta , \epsilon +\hbar\omega+i\eta,\epsilon 
+2\hbar\omega+i2\eta)
- (\eta \to -\eta) \right \} 
\nonumber\\&
+\int^{+\infty}_{-\infty} \frac{d\epsilon}{2\pi i} n_{\rm F}(\epsilon) \left \{ 
P(\epsilon-\hbar\omega-\hbar\omega_2-i2\eta,\epsilon-\hbar\omega-i\eta, \epsilon+i\eta,\epsilon+\hbar\omega+i\eta)
- (\eta \to -\eta) \right \} 
\nonumber\\& 
+\int^{+\infty}_{-\infty} \frac{d\epsilon}{2\pi i} n_{\rm F}(\epsilon) \left \{  
P(\epsilon-3\hbar\omega-i3\eta,\epsilon-2\hbar\omega-i2\eta, \epsilon
-\hbar\omega-i\eta,\epsilon+i\eta)
- (\eta \to -\eta) \right \}~. 
\end{align}
Since $\eta \to 0^+$, we have 
\begin{align}
 B(\omega) &=
\int^{+\infty}_{-\infty} \frac{d\epsilon}{2\pi i} n_{\rm F}(\epsilon)
 \Big \{ 
P^{\rm RRRR}(\epsilon,\epsilon+\hbar\omega, \epsilon+2\hbar\omega,\epsilon+
3\hbar\omega)
\nonumber\\&-
P^{\rm ARRR}(\epsilon,\epsilon+\hbar\omega, \epsilon+2\hbar\omega,\epsilon
+3\hbar\omega)
\Big \} 
\nonumber\\&
+ \int^{+\infty}_{-\infty} \frac{d\epsilon}{2\pi i} n_{\rm F}(\epsilon)
\Big\{  
P^{\rm ARRR}(\epsilon-\hbar\omega ,\epsilon  , \epsilon +\hbar\omega ,\epsilon +2\hbar\omega )
-
P^{\rm AARR}(\epsilon-\hbar\omega ,\epsilon  , \epsilon +\hbar\omega ,\epsilon +2\hbar\omega )
 \Big \} 
\nonumber\\&
+\int^{+\infty}_{-\infty} \frac{d\epsilon}{2\pi i} n_{\rm F}(\epsilon) 
\Big \{ 
P^{\rm AARR}(\epsilon-2\hbar\omega ,\epsilon-\hbar\omega , \epsilon,\epsilon+\hbar\omega )
-
P^{\rm AAAR}(\epsilon-2\hbar\omega ,\epsilon-\hbar\omega , \epsilon,\epsilon+\hbar\omega )
\Big \} 
\nonumber\\& 
+\int^{+\infty}_{-\infty} \frac{d\epsilon}{2\pi i} n_{\rm F}(\epsilon) 
\Big \{  
P^{\rm AAAR}(\epsilon-3\hbar\omega ,\epsilon-2\hbar\omega , \epsilon-\hbar\omega ,\epsilon )
\nonumber\\&-
P^{\rm AAAA}(\epsilon-3\hbar\omega,\epsilon-2\hbar\omega, \epsilon-\hbar\omega,\epsilon)
\Big \}~. 
\end{align}
Note that ``R'' and ``A'' superscript stand for the retarded and advanced, respectively. We shift $\epsilon$ in such a way that all $P$-function arguments is $(\epsilon,\epsilon+\hbar\omega, \epsilon
+2\hbar\omega,\epsilon+3\hbar\omega)$: 
\begin{align}
 B(\omega) &=
\int^{+\infty}_{-\infty} \frac{d\epsilon}{2\pi i} n_{\rm F}(\epsilon)
 \Big \{ 
P^{\rm RRRR}(\epsilon,\epsilon+\hbar\omega, \epsilon+2\hbar\omega,\epsilon
+3\hbar\omega)
\nonumber\\&-
P^{\rm ARRR}(\epsilon,\epsilon+\hbar\omega, \epsilon+2\hbar\omega,\epsilon
+3\hbar\omega)
\Big \} 
\nonumber\\&
+ \int^{+\infty}_{-\infty} \frac{d\epsilon}{2\pi i} n_{\rm F}(\epsilon+\hbar\omega)
\Big\{  
P^{\rm ARRR}(\epsilon ,\epsilon +\hbar\omega , \epsilon+2\hbar\omega ,\epsilon+
3\hbar\omega )
\nonumber\\&-
P^{\rm AARR}(\epsilon ,\epsilon +\hbar\omega , \epsilon +2\hbar\omega ,\epsilon 
+3\hbar\omega )
 \Big \} 
\nonumber\\&
+\int^{+\infty}_{-\infty} \frac{d\epsilon}{2\pi i} n_{\rm F}(\epsilon+\hbar\omega_1+\hbar\omega_2) 
\Big \{ 
P^{\rm AARR}(\epsilon ,\epsilon+\hbar\omega , \epsilon+2\hbar\omega,\epsilon
+3\hbar\omega )
\nonumber\\&-
P^{\rm AAAR}(\epsilon ,\epsilon+\hbar\omega , \epsilon+2\hbar\omega,\epsilon
+3\hbar\omega )
\Big \} 
\nonumber\\& 
+\int^{+\infty}_{-\infty} \frac{d\epsilon}{2\pi i} n_{\rm F}(\epsilon+3\hbar\omega) 
\Big \{  
P^{\rm AAAR}(\epsilon ,\epsilon+\hbar\omega , \epsilon+2\hbar\omega,\epsilon
+3\hbar\omega )
\nonumber\\&-
P^{\rm AAAA}(\epsilon,\epsilon+\hbar\omega, 
\epsilon+2\hbar\omega,\epsilon+3\hbar\omega)
\Big \}~.
\end{align}
Eventually, we obtain
\begin{align}
 B(\omega) &=
 \int^{+\infty}_{-\infty} \frac{d\epsilon}{2\pi i} 
\Big\{ \left [n_{\rm F}(\epsilon) P^{\rm RRRR}-n_{\rm F}(\epsilon+3\hbar\omega) P^{\rm AAAA} \right ] 
+  (n_{\rm F}(\epsilon+\hbar\omega) -n_{\rm F}(\epsilon)  )  P^{\rm ARRR}
\nonumber\\&
+  (n_{\rm F}(\epsilon+2\hbar\omega) - n_{\rm F}(\epsilon+\hbar\omega) )P^{\rm AARR}
+ (n_{\rm F}(\epsilon+3\hbar\omega) -n_{\rm F}(\epsilon+2\hbar\omega) ) P^{\rm AAAR}
\Big\}~.
\end{align}
 \section{Numerical evaluation of the nonlinear optical and dc conductivities}
\begin{figure}[h]
 \includegraphics[width=90mm]{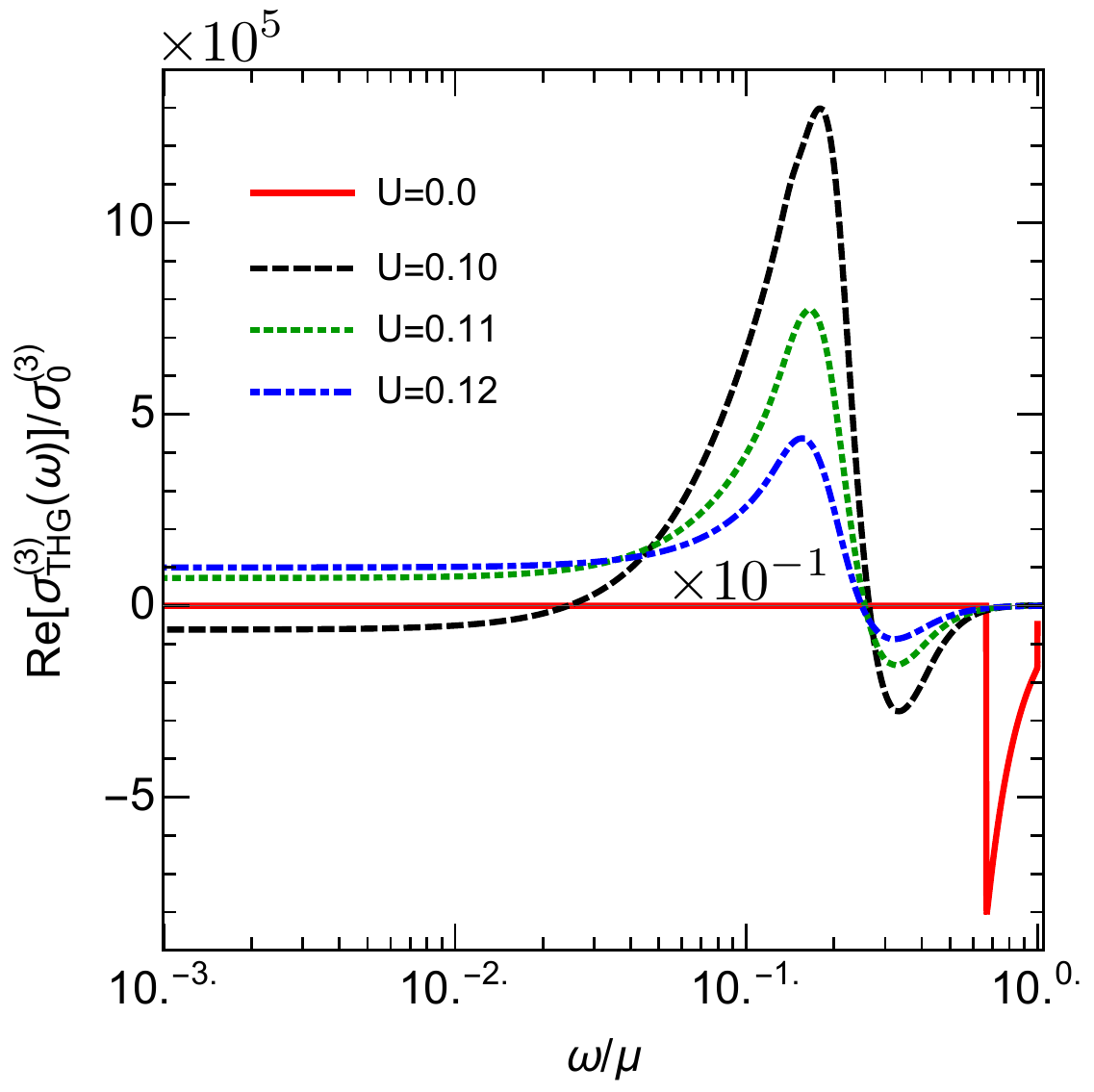}
\caption{Real part of third-harmonic optical conductivity versus frequency in comparison with the non-interacting result. Note that the chemical potential is set $\mu=17~{\rm THz}$, and $\sigma^{(3)}_0 = \sigma_0/E^2_0$.  }
\label{fig:thg}
\end{figure}

In Fig.~\ref{fig:thg}, we plot the frequency dependence of the third-harmonic generation (THG) response function $\sigma^{(3)}_{\rm THG}(\omega)$ for $\mu=17~{\rm THz}$ and few representative values of $U$. For comparison we show also the THG optical response $\sigma^{(3)}_{\rm THG}(\omega)$ for non-interacting electrons which is null for $\hbar\omega \le 2\mu/3$. Note the change of sign of  $\sigma^{(3)}_{\rm dc}$ varying the scattering strength.

\begin{figure}[t]
\includegraphics[width=180mm]{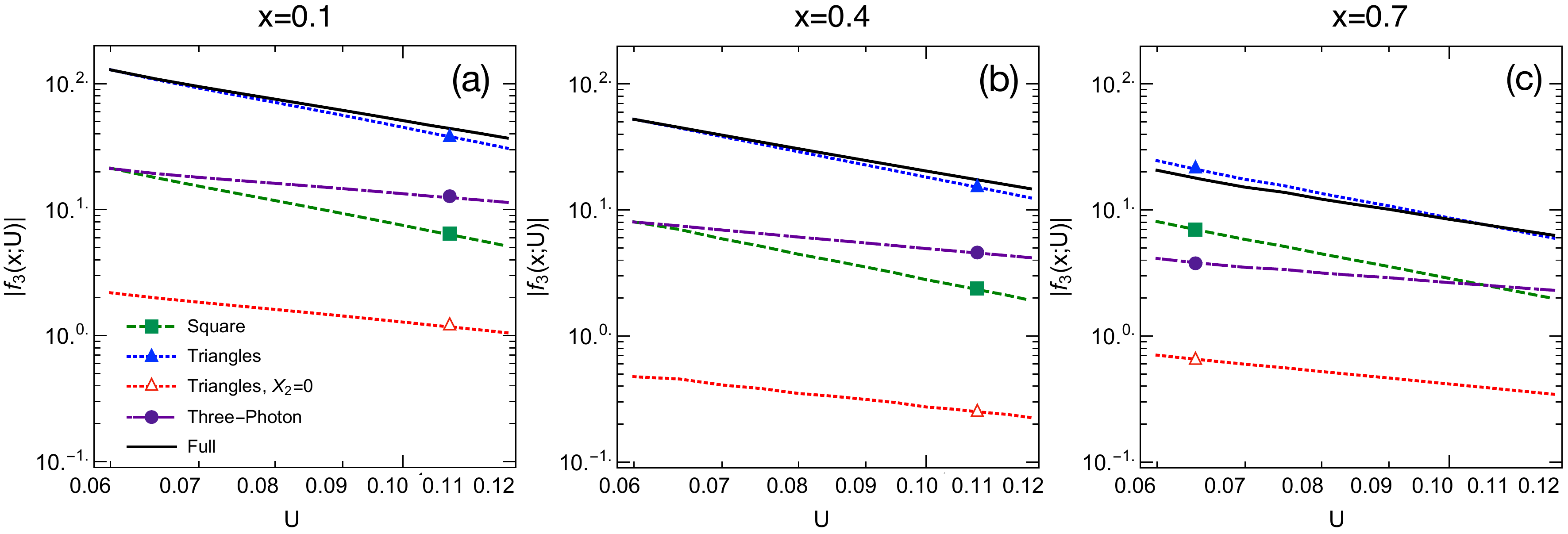}
\caption{Log-log scale plot for the absolute value of the universal $f_3(\mu,U)$ function versus $U$ at $x =0.1,0.4$ and 0.7 which are respectively depicted in panel a, b and c. 
Different lines correspond to the individual contribution of diagrams in Fig.~1a of the main text. }
\label{fig:scaling}
\end{figure}

In Fig.~\ref{fig:scaling}, we illustrate the universal scaling of $f_3$ versus $U$ in the quantum regime for different values of $x=\mu/\Gamma(\mu)<1$. As seen the slop of the curves in the log-log scale plot does not strongly depends on the value of $x$ which support the validity of Eq.~(5) given in the main text. 

\begin{figure}[t]
\includegraphics[width=100mm]{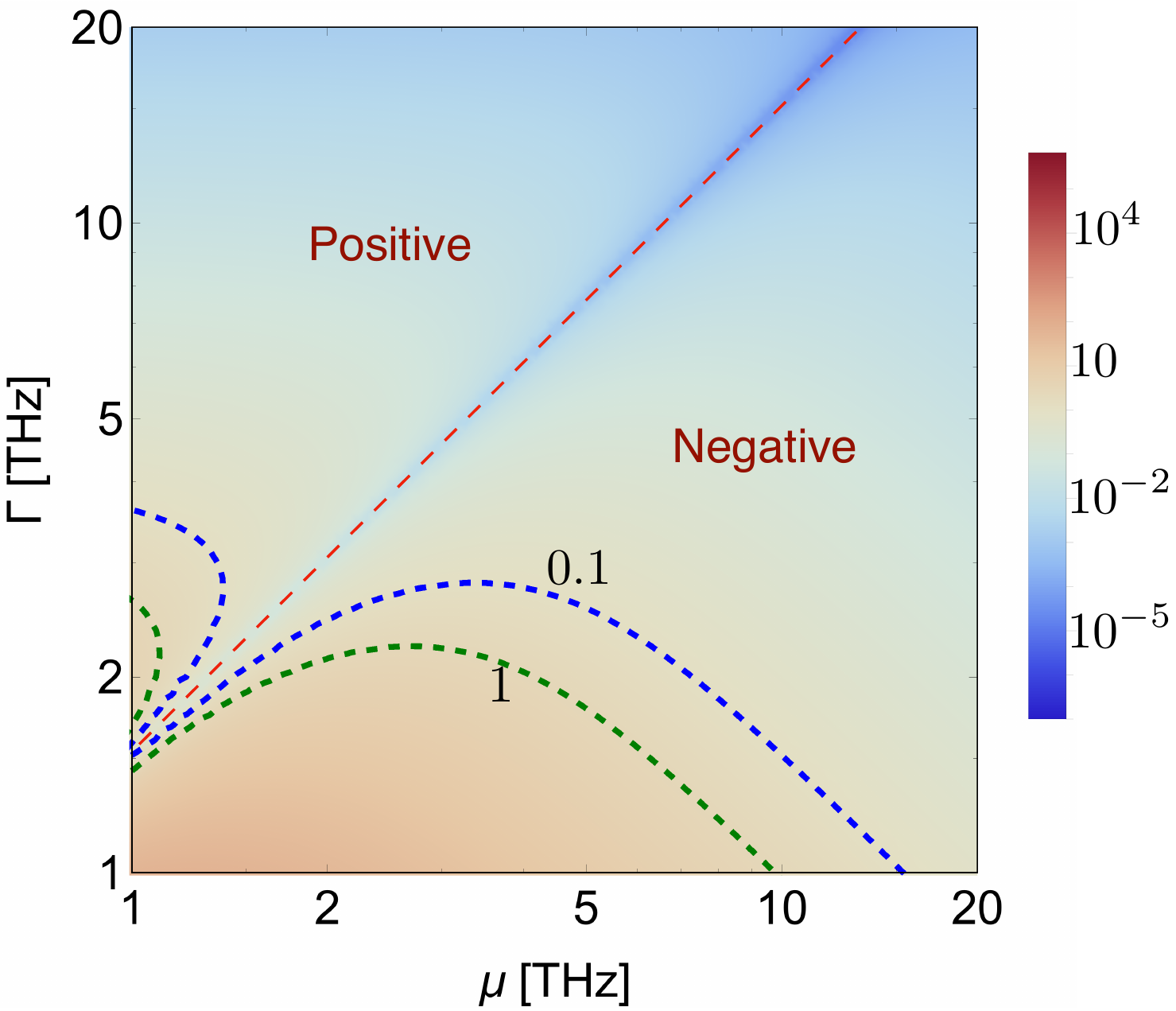}
\caption{Colormap plot for $ g = |\sigma^{(3)}_{\rm dc} E^2/\sigma^{(1)}_{\rm dc}|$ factor with $ E=1{\rm mV/nm}$ versus chemical potential $\mu$ and scattering rate $\Gamma$ in the constant-$\Gamma$ model, $\Sigma = - i\Gamma$. The sign of $\sigma^{(3)}_{\rm dc}$ is written on the plot where the sign-switch border is highlighted by a dashed red line. Green and blue dotted lines stand for the contour lines with $g=1$ and $g=0.1$, respectively.}
\label{fig:constant_gamma}
\end{figure}

In Fig.~\ref{fig:constant_gamma}, we show the phase diagram for the constant-$\Gamma$ model. As it is seen this phase diagram is completely different from that of the full quantum theory which is given in Fig.~3c of the main. text. We can see only one sign-change in the constant-$\Gamma$ model in contrast to that of full quantum theory which gives two sing-changes. Unlike the full quantum theory, the constant-$\Gamma$ model predicts a positive nonlinear correction in the quantum regime.  

 \end{document}